\documentclass[acmlarge]{acmart}
\AtBeginDocument{%
  \providecommand\BibTeX{{%
    \normalfont B\kern-0.5em{\scshape i\kern-0.25em b}\kern-0.8em\TeX}}}

\copyrightyear{2018}
\acmYear{2018}
\acmDOI{XXXXXXX.XXXXXXX}

\acmConference[Conference acronym 'XX]{Make sure to enter the correct
  conference title from your rights confirmation emai}{June 03--05,
  2018}{Woodstock, NY}
\acmISBN{978-1-4503-XXXX-X/18/06}

\usepackage{graphicx} % 用来插入图片
\usepackage{subcaption} % 用来创建子图
\usepackage{multirow}

\usepackage[utf8]{inputenc}
\usepackage[T1]{fontenc}
\usepackage{hyperref}
\usepackage{url}
\usepackage{booktabs}
\usepackage{amsfonts}
\usepackage{nicefrac}
\usepackage{microtype}
\usepackage{todonotes}
\usepackage{listings}
\usepackage{algorithm}
\usepackage{algpseudocode}
\usepackage{soul}
\usepackage{tabularx}
\usepackage{pifont}
\usepackage{xcolor}
\usepackage{graphicx}
\usepackage{subcaption}
\usepackage[normalem]{ulem}

\newcommand{\greencheck}{\textcolor{green}{\ding{51}}}
\newcommand{\redcross}{\textcolor{red}{\ding{55}}}

\begin{document}

\title[AUGlasses: Continuous Action Unit based Facial Reconstruction ...]{AUGlasses: Continuous Action Unit based Facial Reconstruction with Low-power IMUs on Smart Glasses}

\author{Yanrong Li}
\orcid{0000-0002-7278-9317}
\email{kidominox@gmail.com}
\affiliation{%
  \institution{Institute of Computing Technology, CAS and UCAS}
  \country{China}
}

\author{Tengxiang Zhang}
\authornote{Tengxiang Zhang and Yiqiang Chen are also with Beijing Key Lab. of Mobile Computing and Pervasive Device.}
\authornote{Corresponding Author}
\orcid{0000-0002-0949-2801}
\affiliation{%
  \institution{Institute of Computing Technology, CAS and UCAS}
  \country{China}}
\email{ztxseuthu@gmail.com}

\author{Xin Zeng}
\orcid{0009-0004-9862-4168}
\affiliation{%
  \institution{Institute of Computing Technology, CAS and UCAS}
  \country{China}
}

\author{Yuntao Wang}
\orcid{0000-0002-4249-8893}
\affiliation{%
 \institution{Tsinghua University}
 \country{China}}

\author{Haotian Zhang}
\author{Yiqiang Chen}
\authornotemark[1]
\orcid{0000-0002-8407-0780}
\affiliation{%
  \institution{Institute of Computing Technology, CAS and UCAS}
  \country{China}}

\renewcommand{\shortauthors}{Li et al.}
\newcommand*{\eg}{\textit{e.g.},\;}
\newcommand*{\ie}{\textit{i.e.},\;}
\newcommand*{\vs}{\textit{v.s.}\;}
\newcommand*{\etc}{\textit{etc.}}
\newcommand*{\etal}{\textit{et~al.}\;}
\newcommand\TX[1]{\textcolor{blue}{\textit{\textbf{TX}: #1}}}

\newcommand{\code}[1]{\texttt{#1}}
\begin{abstract}
Recent advancements in augmented reality (AR) have enabled the use of various sensors on smart glasses for applications like facial reconstruction, which is vital to improve AR experiences for virtual social activities. 
However, the size and power constraints of smart glasses demand a miniature and low-power sensing solution. 
AUGlasses achieves unobtrusive low-power facial reconstruction by placing inertial measurement units (IMU) against the temporal area on the face to capture the skin deformations, which are caused by facial muscle movements. 
These IMU signals, along with historical data on facial action units (AUs), are processed by a transformer-based deep learning model to estimate AU intensities in real-time, which are then used for facial reconstruction. 
Our results show that AUGlasses accurately predicts the strength (0-5 scale) of 14 key AUs with a cross-user mean absolute error (MAE) of 0.187 (STD = 0.025) and achieves facial reconstruction with a cross-user MAE of 1.93 mm (STD = 0.353). 
We also integrated various preprocessing and training techniques to ensure robust performance for continuous sensing. 
Micro-benchmark tests indicate that our system consistently performs accurate continuous facial reconstruction with a fine-tuned cross-user model, achieving an AU MAE of 0.35.
\end{abstract}




\begin{teaserfigure}
  \centering
  \includegraphics[width=0.9\textwidth]{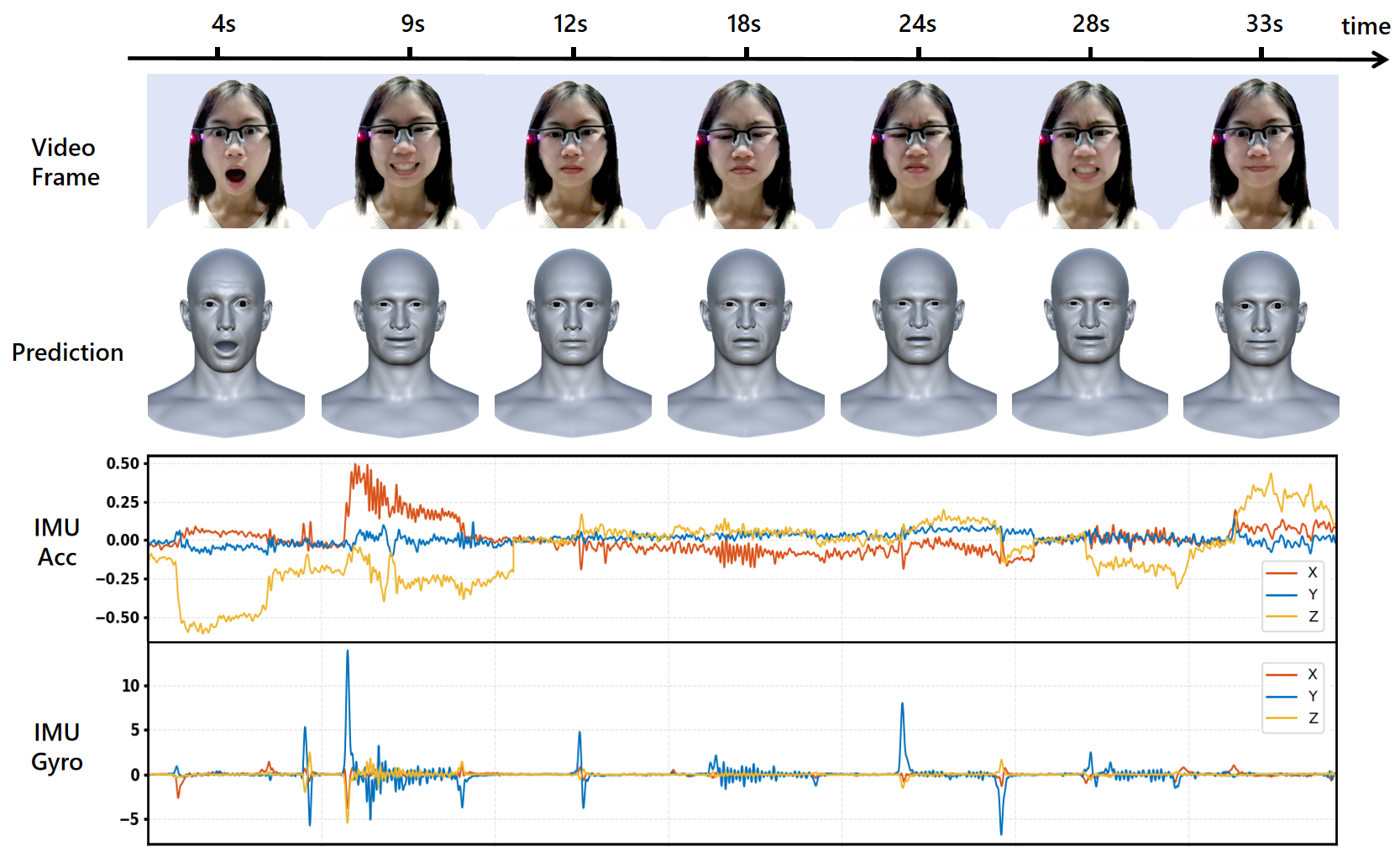}
  \caption{Demonstration of continuous 3D facial reconstruction of AUGlasses.}
  % \Description{Enjoying the baseball game from the third-base
  % seats. Ichiro Suzuki preparing to bat.}
  \label{fig:teaser}
\end{teaserfigure}


\maketitle

\section{Introduction}

%% facial movement recognition or facial action recognition?
Facial reconstruction enables a plethora of applications since
human faces contain rich information regarding emotional states~\cite{bassili_emotion_1979, krumhuber_role_2023}, health states~\cite{argaud_facial_2018,roter_expression_2006}, dietary activities~\cite{bedri_fitbyte_2020, bedri_earbit_2017, bi_auracle_2018}, and even fatigue level~\cite{uchida_identification_2018}.  
Users can move facial muscles for interaction~\cite{zhang_facilitating_2019} and active feedback purposes~\cite{yan_frownonerror_2020}. 
It is also key for realistic digital avatars of humans, which is important for metaverse applications like virtual conference~\cite{frith_role_2009, cha_real-time_2020}, swiftly establishing understanding and connection with others in the virtual space. 

Researchers have used various types of sensors to collect data for replicating facial expressions and muscle movements in the digital realm.
Cameras~\cite{valle_face_2019, dong_style_2018} and millimeter-wave radars~\cite{zhang_mmfer_2023} are capable of capturing facial features comprehensively. However, their dependence on third-person view angles poses challenges in environments where such perspectives are difficult to maintain.
Although it is feasible to infer characteristics of the entire face using only local facial regions, this approach often requires significant power consumption. Notably, previous researches have incorporated sensors into various wearable items such as necklaces~\cite{chen_neckface_2021}, headphones~\cite{choi_ppgface_2022, song_facelistener_2022}, hats~\cite{bello_inmyface_2023}, ear-mounted devices~\cite{bedri_earbit_2017, bi_auracle_2018}, and eyeglasses~\cite{xie_acoustic-based_2021} to detect facial movements without direct full-face capture. However, the power requirements of these sensors, including RGB cameras and acoustic sensors like microphones and speakers~\cite{xie_acoustic-based_2021, song_facelistener_2022}, are typically too high for continuous operation, limiting their practicality for prolonged use.

%%In fact, it is possible to infer features of the entire face by collecting information only from local facial areas. 
%%For example, previous research integrates sensors into necklaces\cite{chen_neckface_2021}, headphones\cite{choi_ppgface_2022, song_facelistener_2022}, hats\cite{bello_inmyface_2023}, ear-mounted device\cite{bedri_earbit_2017,bi_auracle_2018}, and eyeglasses\cite{xie_acoustic-based_2021} to recognize facial movements without directly capturing the entire face. 
%%However, the above-mentioned sensors consume plenty of power, which greatly impacts the working time of wearables. 
%%For example, both RGB camera and acoustic sensors like microphones and speakers~\cite{xie_acoustic-based_2021, song_facelistener_2022} consume around 150mA at working state, which will drain a 350mAh lithium polymer battery in three hours. 

In this paper, we present AUGlasses, a pair of smart glasses equipped with three low-power inertial measurement units (IMUs) for precise facial reconstruction. 
These IMUs have a high sensitivity for rapid and accurate capture of subtle and evanescent skin deformations caused by facial muscle movements. 
%We conducted a systematic analysis and found that skin deformation is most pronounced near the temples in the upper facial region, a finding corroborated by facial anatomy. 
We conducted a systematic analysis and found that skin deformation is most pronounced near the temporalis muscle above the zygomatic arch, a finding corroborated by facial anatomy. 
To exploit this, we have incorporated a bio-compatible, elastic structure in the glasses, positioning two IMUs against the temporal area to monitor skin movements and shape changes.
The data from these IMUs, combined with historical records of facial action unit (AU) intensities, feed into a deep learning model designed to predict current AU intensity. Opting to predict AU intensities rather than facial feature points enhances the system's generalizability. The result is a high-frequency AU sequence output, operating at 30 frames per second, which ensures real-time responsiveness and utility for downstream applications.
%The IMU signals, along with historical AU strengths, are then fed into a deep learning model to estimate the current strengths of facial action units (AUs).
%We choose to estimate AU strengths instead of facial feature points to maximize the generalizability of our system. The output of our system is a high-frequency AU sequence, designed to operate at 30 frames per second, providing real-time responsiveness and work for downstream tasks.

A new training strategy called Prefix-conditioned Sequence Forecasting was proposed to force the model to learn long-range dependency exhibited across many high-frequency frames.% A novel transformer based model incorporates both the IMU signals and previous estimated AUs to predict future Au strengths. 
The estimated AU strengths are then fed into Unity for real-time facial reconstruction. 

The advantages of our system lie in three aspects:
First, our low-power sensing method ensures minimal impact on the smart glasses' working lifetime. 
The power consumption of the IMU and Bluetooth communication is only 49.95 mW (13.5 mA at 3.7 V), lower than current solutions. 
%####
If the sensor only works 25\% of the time, the average power consumption is as low as 12.5 mW. 
It is also privacy-preserving, lightweight, and comfortable.
IMUs will not expose sensitive personal data like images or voices, which eliminates user privacy concerns. 
The supporting elastic gel structure and the IMUs only weigh 0.83 grams in total, which is only 1.66\% of current commercial smart glasses (\ie 50g\footnote{\url{https://www.meta.com/smart-glasses/wayfarer-matte-black-graphite-polar-gradient/}}). % ####
Even though the IMUs are in contact with the skin, the contact point is also actually small.
So the additional weight and skin contact introduced by our sensors are barely noticeable to users. 
Second, we have devised a series of techniques to ensure the accuracy of long-term continuous facial reconstruction. 
In the pre-processing phase, we updated mapping parameters for moving artifact removal to ensure its effectiveness as the user moves around;
in the inference phase, we reset the values of AUs are the model input to avoid prediction error propagation during long-time sensing; in the training phase, we designed a novel prefix-conditioned sequence forecasting strategy to enforce the model learn from a prefixed sequence of observed data, allowing it to forecast subsequent sequences based on a broader context rather than just the immediate past. 
This avoids exposure bias and improves the accuracy and reliability of continuous facial reconstruction.
Third, our system reconstructs the entire face by predicting only 14 AUs.
The number of predicted AUs is limited by our sensing approach, which is detecting the skin movement of one spot on the face using IMUs.
However, as we will show later, the facial reconstruction result is surprisingly accurate even for continuous sensing.
This is enabled by inter-dependencies of different facial muscles as they pull or suppress one another. 
By limiting the output AUs, we avoid the complexity introduced by directly predicting 2D coordinates of feature points on the face while achieving great facial reconstruction performance at the same time. 

We first conducted a series of experiments to optimize the sensor location and the parameters of the supporting structure, then conducted user studies for both sitting and walking scenarios to evaluate the prediction accuracies of 14 AU strengths, the facial reconstruction and long-time continuous forecasting. 
Our user study results show that AUGlasses can accurately predict the strengths (0-5) of 14 AUs with a within-user (cross-session) MAE of 0.21 (STD = 0.028) and cross-user MAE of 0.19 (STD = 0.025).   
Based on the estimated AU strengths, our system achieves accurate and robust facial reconstruction with a within-user (cross-session) MAE of 1.89mm (STD=0.35) and cross-user MAE of 1.93 mm (STD = 0.353).
Fig~\ref{fig:teaser} shows the exemplary facial reconstruction results corresponds with such MAEs for a straightforward demonstration of the reconstruction performance.
The MAE only deteriorates to 0.4 and 0.5 within 30 seconds of the estimation time, validating our system's ability for continuous facial reconstruction. 

The contributions of this article are:
\begin{itemize}
\item We proposed a novel low-power and lightweight facial reconstruction method on smart glasses by sensing the skin movements around the temporal area using two IMUs. 
\item We designed a signal processing pipeline to ensure accurate facial reconstruction. Specifically, we developed various techniques in signal pre-processing and model training stages to ensure accurate prediction results for long-time continuous sensing. 
\item We evaluated our prototype comprehensively to show the feasibility and usability of IMU-based facial reconstruction under different test scenarios, which is valuable for future IMU-based skin deformation sensing work. 
\end{itemize}

\section{Related Work} 

\begin{table}[t]
    \centering
    \caption{Relevant facial sensing work with head wearable devices}
    \label{table:related_work}
    \resizebox{\columnwidth}{!}{
    \begin{tabular}{lcccccc}
        \toprule
        \textbf{Name} &
        \textbf{Device} &
        \textbf{Signal} &
        \textbf{Unobtrusive} &
        \textbf{Privacy-preserving} &
        \textbf{Low-power} &
        \textbf{Facial Reconstruction} \\
        \midrule
        C-Face \citep{chen_c-face_2020} & Ear-mounted & Camera & \redcross & \redcross & \redcross  & \redcross \\
        \hline
        NeckFace \citep{chen_neckface_2021} & Neck-mounted & IR cameras & \redcross & \greencheck & \redcross & \greencheck \\
        \hline
        FaceRecGlasses \citep{aoki_facerecglasses_2021} & Glasses & Camera & \greencheck & \redcross & \redcross & \redcross \\
        \hline
        TexonMask \citep{guo_texonmask_2023} & Mask & Textile Electrodes & \redcross & \greencheck & \greencheck & \redcross \\
        \hline
        EARFace \citep{zhang_i_2023} & Ear-mounted & Ultrasonic & \greencheck & \greencheck & \redcross & \greencheck\\
        \hline
        EarIO \citep{li_eario_2022} & Ear-mounted & Microphone and speaker & \greencheck & \redcross & \greencheck & \greencheck\\
        \hline
        ExpressEar \citep{verma_expressear_2021} & Ear-mounted & IMU-Only & \greencheck & \greencheck & \redcross & \redcross\\
        \hline
        \textbf{AUGlasses}(Ours) & Glasses & IMU-Only & \greencheck & \greencheck & \greencheck & \greencheck\\
        \bottomrule
    \end{tabular}
    }
\end{table}

In this paper, we propose and evaluate the performance of AUGlasses in facial reconstruction by atomic facial expressions using IMU-augmented glasses. 
The related work can be divided into three main areas: various approaches for facial reconstruction, advancements in smart eyewear computing, and state-of-the-art learning techniques for time series signal-based recognition.

\subsection{Facial Reconstruction}

Facial reconstruction is a critical research area with significant implications across a variety of applications within Human-Computer Interaction, such as augmented reality (AR)~\cite{zhang_dense_2018}, virtual reality (VR)~\cite{liang_new_2004}, accessibility technologies~\cite{kim_development_2006}, medical and healthcare applications~\cite{mao_facial_2010}, and the entertainment industry~\cite{danieau_automatic_2019}. 
The most common method for facial reconstruction uses the visual modality, leveraging facial images and computer vision techniques to reconstruct facial features comprehensively. Work in this area has utilized diverse data sources, including extensive photo collections of the same subject~\cite{roth_adaptive_nodate, roth_unconstrained_2015}, sequences of video frames~\cite{weise_faceoff_2009, cao_real-time_2015}, and even rough depth maps or single images~\cite{cao_real-time_2015, richardson_learning_2017}. Devices like C-Face~\cite{chen_c-face_2020}, NeckFace~\cite{chen_neckface_2021}, and FaceRecGlasses~\cite{aoki_facerecglasses_2021} represent significant advancements in wearable vision technology, designed to facilitate real-life facial reconstruction with minimal intrusion.
Despite the achievements with traditional visual modalities, challenges related to lighting, privacy, and energy efficiency have prompted research into other sensor technologies for facial reconstruction. 

The shift from visual to other modalities has introduced novel approaches utilizing proximity sensors, electromyography (EMG), pressure sensors, electrooculography (EOG), and audio signals to monitor facial expressions and muscle movements. Innovations in this space include EarFS~\cite{matthies_earfieldsensing_2017}, which uses electrodes in the ear canal to detect facial gestures, and TexonMask~\cite{guo_texonmask_2023}, which integrates textile electrodes in facemasks for expression recognition.
Systems such as EARFace~\cite{zhang_i_2023} and EarIO~\cite{li_eario_2022} demonstrate the use of acoustic sensors in earphones to track facial movements, offering privacy-preserving solutions. Additionally, newer technologies like millimeter wave radar have been explored for their potential to perform non-intrusive and privacy-preserving facial reconstructions, as seen in studies like mmFER~\cite{zhang_mmfer_2023} and mm3DFace~\cite{xie_mm3dface_2023}.

Despite the advancements in sensor technologies for facial reconstruction, challenges in granularity, privacy, and energy efficiency persist, highlighting the need for a modality that can offer detailed and continuous monitoring without significant drawbacks. Inertial Measurement Units (IMU) provide a promising solution by capturing subtle facial muscle movements through micro-movements, offering a discrete and energy-efficient approach ideal for sensitive applications in medical diagnostics and personal security.

In the realm of inertial sensors, systems like ExpressEar~\cite{verma_expressear_2021} demonstrate the effectiveness of IMUs placed in commercial dual-earpiece earables for discreetly detecting facial expressions. Our research progresses further by employing smart eyewear equipped with IMUs, strategically positioned to capture a broader spectrum of facial movements. This optimal placement significantly enhances the granularity and scope of facial reconstruction, making our approach well-suited for everyday use and extending the capabilities of smart eyewear and wearable computing. AUGlasses are designed with multiple key advantages, including being unobtrusive, preserving privacy, consuming low power, and supporting fine-grained facial action unit forecasting. For a detailed comparison of these features with other systems, please see Table~\ref{table:related_work}.

\subsection{Smart Eyewear}

Smart eyewear has emerged as a crucial wearable device due to its convenience and seamless integration into daily life, making it an ideal platform for HCI applications. These devices typically host a range of sensors and have been used for diverse applications including health monitoring, emotional state detection, and even environment context sensing~\cite{zhang_bleselect_2023}.

The exploration of smart eyewear for facial sensing is not new. Devices like Expression Glasses have been designed to detect facial muscle movements using piezoelectric sensors, achieving initial high accuracy levels in expression recognition~\cite{scheirer_expression_1999}. Similarly, JINS MEME eyeglasses incorporate sensors to monitor eye movements and head motion, evaluating user concentration~\cite{uema_jins_2017}. Further advancements include the development of smart eyewear equipped with photo-reflective sensors or proximity sensors, used for applications ranging from facial expression recognition~\cite{masai_evaluation_2017} to dietary monitoring~\cite{saphala_proximity-based_2022}.
Despite these innovations, eyewear-based facial sensing technologies often face challenges such as coarse granularity in data capture and high power consumption. These limitations restrict their practicality for continuous and detailed monitoring tasks.

Moreover, while there have been substantial studies utilizing various sensors embedded in eyewear, there is a noticeable gap in research specifically using inertial measurement units for the complex task of facial reconstruction. No existing smart eyewear solely employs IMU for detailed and continuous facial expression mapping, a gap our research aims to fill. This novel approach could potentially overcome the granularity and power efficiency challenges faced by current technologies, paving the way for more effective and user-friendly smart eyewear applications.

\subsection{Learning Algorithms for Time Series Signal Based Recognition}
\label{sec:releted_trainging}

The evolution of algorithms in time series signal-based recognition has progressed from classical methods, such as Support Vector Machines (SVM)~\cite{park_online_2012, kwapisz_activity_2011}, to more advanced deep learning approaches that combine convolutional and recurrent neural units~\cite{qian_what_2022, qian_novel_2019, zeng_understanding_2018}. Recently, transformer architectures have demonstrated exceptional performance, leveraging their innate ability to handle sequential data effectively, thus shining in tasks involving signal pattern recognition.

In specific application domains, transformer-based models have been particularly effective. For instance, in emotion recognition, self-supervised learning models using transformers have excelled by capturing relevant features from ECG time-series signals, showcasing superior performance on emotion recognition tasks~\cite{vazquez-rodriguez_transformer-based_2022}. Similarly, for human activity recognition, lightweight transformer approaches have been applied successfully using IR-UWB radar~\cite{li_human_2023} and EEG signals~\cite{wan_eegformer_2023}. Even WiFi signals have been utilized for indoor human mobility modeling through transformer-based techniques~\cite{trivedi_wifimod_2021}.
Moreover, the adaptation of transformers for multivariate time series representation learning has led to notable advancements in tasks requiring regression and classification, demonstrating their effectiveness over traditional methods even with limited training data~\cite{nie_time_2023}. Additionally, innovative frameworks combining convolutional layers with transformers, such as the IF-ConvTransformer, have further enhanced human activity recognition capabilities using IMU fusion~\cite{zhang_if-convtransformer_2022}.

% \begin{figure}[t]
%   \centering
%   \includegraphics[width=\linewidth]{img/related_work_model.pdf}
%   \caption{(a) Training Stage: The traditional autoregressive transformer model quickly learns shortcuts to fit the training data, effectively learning a trick rather than understanding underlying patterns. (b) Prediction Stage: Due to its reliance on superficial learning, the model performs poorly, failing to generalize effectively in real-world applications.}
%   \label{fig:relatedwrok_model}
% \end{figure}

However, directly applying conventional methods to our task is not feasible due to the distinct characteristics of our signals. Firstly, using IMUs to monitor facial muscle movements is inherently susceptible to interference from head movements. Vigorous head movements can overwhelm the subtle signals of facial muscle activities, making them difficult to detect accurately. Secondly, autoregressive architectures like transformers inherently suffer from exposure bias~\cite{liu_confidence-aware_2021, leblond_machine_2021}, a problem exacerbated in our context by the high-frequency nature of our predictions (30 outputs per second). This issue becomes more pronounced with such frequent outputs, severely impacting the reliability of long-term predictions.
These challenges necessitate the development of tailored techniques, including a dedicated preprocessing method for artifact removal and a specific training strategy. We employ a prefix-conditioned prediction method, which replaces the standard auto-regressive time series prediction. This approach is specifically designed to enforce the model to learn long-term dependencies, enabling effective long-duration forecasting without succumbing to the limitations posed by exposure bias and external motion interference. These modifications enhance the overall effectiveness and reliability of our system, making it uniquely suited for continuous and accurate facial movement detection.

% the effectiveness of IMUTube~\cite{On the Effectiveness of Virtual IMU Data for Eating Detection with Wrist Sensors}, a system that generates virtual IMU data from video for human activity recognition, demonstrating significant improvements in detecting subtle activities like eating, thereby expanding the potential applications of virtual sensor data in activity recognition systems. % IMU 虚拟数据生成

% 通过固定输入时的history au block 和 history imu signal block， 我们在保留历史信息的同时，强制模型去学习较为长时间的依赖关系，而避免模型陷入简单的复制上一帧的AU结果。

% Tengxiang: need to be around facial reconstruction. 1. facial reconstruction based on signals other than IMU; 2. facial expression recognition/reconstruction based on IMU; 3. deep learning architecture

\section{Preliminary}
%AU definition - skin deformation based sensing - verified by previous non-contact solutions - contact solution - contact point - experiment - 1. skin deformatoion temple area; 2. IMU signal capturing 
\subsection{Facial Action Unit}
We use action units defined by the Facial Action Coding System (FACS) as our model output for facial reconstruction.
FACS is a comprehensive framework for identifying and categorizing physical expressions of human emotion through facial muscle movements. 
The human face contains dozens of muscles distributed around the eyes, nose, and other facial areas, driving facial movements~\cite{wu_modelling_2013}.
For instance, the orbicularis oculi muscle is mainly responsible for closing the eyelids, and the zygomaticus major muscle is involved in the process of raising the corners of the mouth.
Developed by psychologists Paul Ekman and Wallace V. Friesen in the 1978~\cite{ekman_facial_1978}, FACS breaks down facial expressions into individual components called action units (AUs), each corresponding to a specific underlying facial muscle movement. 
For example, AU1, the Inner Brow Raiser, involves the contraction of the frontalis, and pars medialis muscle. 
This action leads to the raising of the inner portions of the eyebrows, commonly seen in expressions of surprise, concern, or fear, as it helps to widen the eyes. 
Another example is AU12, the Lip Corner Puller, where the zygomaticus major muscle is contracted to pull the lip corners upward and outward. 
This action unit is essential in forming smiles, indicative of emotions such as happiness or contentment, and is also commonly used in social or polite smiling scenarios.
These action units can occur individually or combine to form rich expressions.
This system allows for a detailed, objective, and standardized measurement of facial expressions, independent of interpretation or context, which we use as basis for facial reconstruction.  

\subsection{Skin-deformation based AU Sensing}
There are different ways to capture the facial AU intensity changes.
EMG sensors can record the level of facial muscle activation, which can be directly mapped to AU intensities~\cite{lapatki_surface_2003}.
However, such methods require deploying multiple sensing electrodes on the face with a high power consumption, making it less practical for everyday use. 
Another way to sense muscle activation levels is through monitoring skin shape deformation.  
This works because facial muscles end in superficial fascia, dermis, and even other muscles.
When facial muscles contract, they can push and pull on the surface of the skin, causing the surface of the skin and fat pads to deform.
So instead of directly measuring the muscle stimulation level, many previous works detect AUs by sensing skin deformations caused by facial muscle activation and movements. 
For instance, acoustic transmitters and receivers\cite{li_eario_2022} have been used for facial reconstructions since skin deformations cause different signal reflective patterns for different facial expressions. 

Even though sensors like speakers and microphones do not directly contact the skin, they are usually heavy and consume more power due to their active sensing nature, which deteriorates the usage experience. 
Contact-based sensing solutions usually consume less power without actively sending out signals.
For example, our system places low-power IMUs against the skin so that it moves as the skin deforms, effectively acquiring facial muscle movement information.  
However, sensors deployed directly on the face may cause uncomfortableness and raise aesthetic concerns. 

\subsection{Sensor Placement Optimization}
One of the advantages of skin-deformation-based AU sensing is that movements of different muscles will cause deformations at the same spot on the face, so there is no need to place one IMU for each facial muscle. 
However, it is not clear where is the optimal spot to place IMU so that it can be easily integrated into IMUs and detect as many AUs as possible. 
So we conduct an experiment to optimize IMU placement locations on the face, optimizing for the most significant amplitude variations in facial muscle movements across different AU expressions.

\begin{figure}[t]
    \centering
    \captionsetup[subfigure]{justification=centering} % 设置子图的标题为居中
    \begin{subfigure}[b]{0.245\linewidth}
        \includegraphics[width=0.9\linewidth,height=3cm]{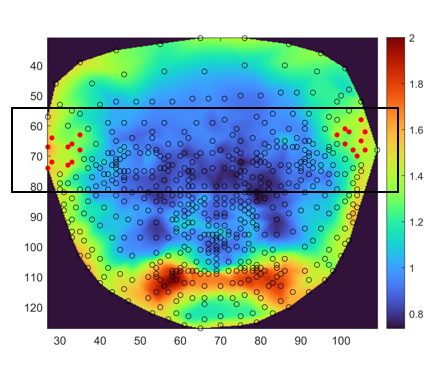}
        \caption{Heat map of changes in facial muscles}
        \label{fig:hotmap}
    \end{subfigure}
    % \hfill % 在子图之间添加空格
    \begin{subfigure}[b]{0.245\linewidth}
        \includegraphics[width=0.9\linewidth,height=3cm]{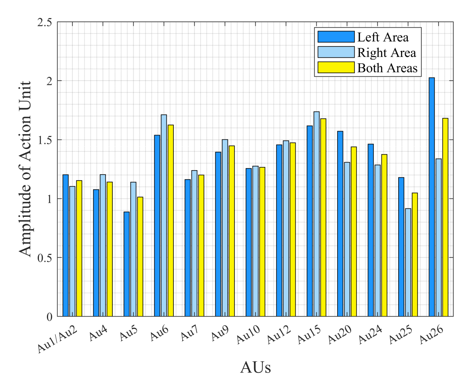}
        \caption{Amplitude of different action units}
        \label{fig:muscalebar}
    \end{subfigure}
    \begin{subfigure}[b]{0.245\linewidth}
        \includegraphics[width=0.9\linewidth,height=3cm]{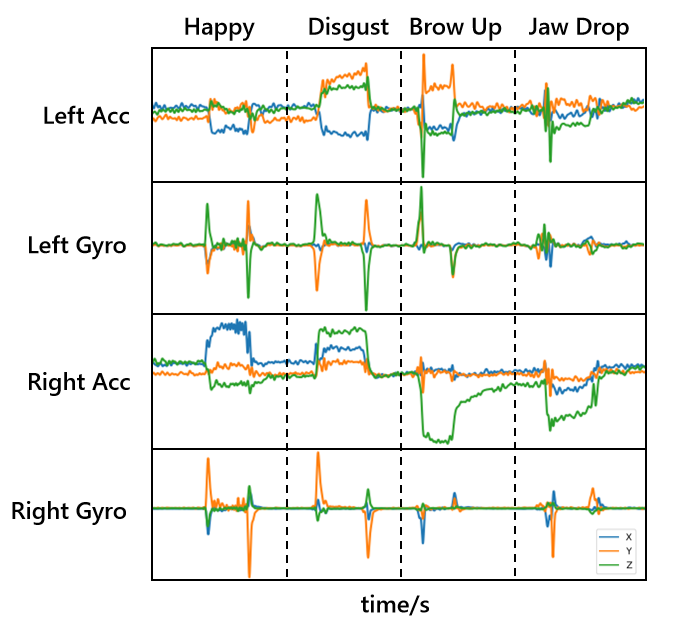}
        \caption{IMU signals of different facial movements.}
        \label{fig:ex_imu}
    \end{subfigure}
    % \hfill % 在子图之间添加空格
    \begin{subfigure}[b]{0.245\linewidth}
        \includegraphics[width=0.9\linewidth,height=3cm]{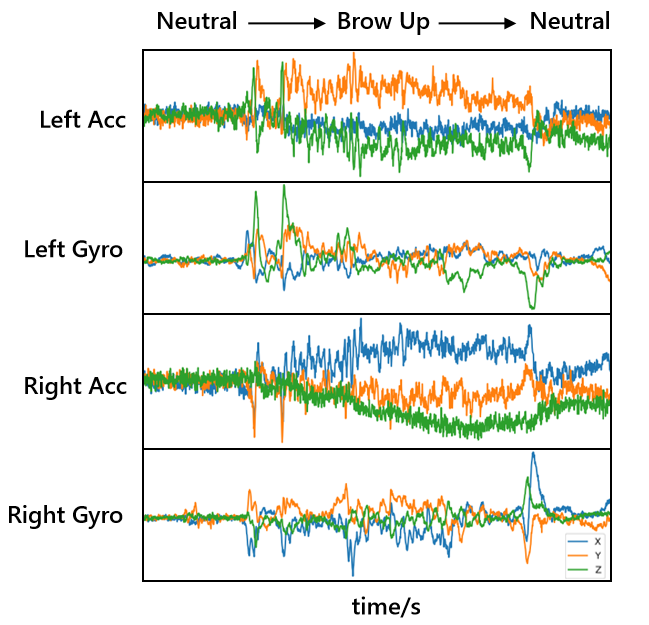}
        \caption{IMU signal of slow expression changes.}
        \label{fig:browup}
    \end{subfigure}

    \caption{Exploring changes in facial muscles within the scope of eyeglasses.}
    \label{fig:facialmuscle}
\end{figure}

We picked 14 AUs for this pilot study and later user studies for facial reconstruction.
On one hand, we need to ensure the selected AUs can be predicted by OpenFace, which we use for ground-truth collection; on the other hand, the AUs should have strong intensities when making the 7 typical facial expressions (\ie sadness, anger, happiness, surprise, fear, disgust, and neutral). 
We also merged Inner Brow Raiser(AU1) and Outer Brow Raiser(AU2) so that it is easier for participants to perform. 
So the AUs we use for all following experiments are: AU1\/AU2 known as Brow Raiser, AU4 as Brow Lowerer, AU5 as Upper Lid Raiser, AU6 as Cheek Raiser, AU7 as Lid Tightener, AU9 as Nose Wrinkler, AU10 as Upper Lip Raiser, AU12 as Lip Corner Puller, AU15 as Lip Corner Depressor, AU20 as Lip Stretcher, AU24 as Lip Pressor, AU25 as Lips Part, and AU26 as Jaw Drop.

We recruited 4 participants (3 females and 1 males, age ranging from 21 to 50).
We used the LeTMC-520 depth camera \footnote{\url{https://orbbec3d.com/}} for data acquisition. 
We made sure the participants' faces were not obscured, then asked them to sit in front of the camera, 40cm away from the depth camera.
After reviewing the 14 AUs~\cite{zarins_anatomy_2018}, participants kept their heads still and performed each AU in 10 rounds, with a 5-second rest after each action and a 30-second rest after each round.
%During this phase, we only need to capture two frames of participants expressing neutral expressions and AU, rather than the entire video.

To calculate the spatial variations of different facial spots, we used the DAD-3dheads algorithm\cite{martyniuk_dad-3dheads_2022} to extract 565 2D face landmark points from the RGB image and simultaneously retrieve the depth information of these points from the corresponding depth image.
We calculate the movement amplitude $D$ resulting from muscle movements,

\begin{equation}
D = \sqrt{(x_1 - x_2)^2 + (y_1 - y_2)^2 + \varDelta d^2}
\end{equation}

where $x_1$ and $x_2$ represent the x-axis pixel distance in the neutral state and the AU state respectively, $y_1$ and $y_2$ respectively represent the y-axis pixel distance in the corresponding state. $\Delta d$ is the pixel depth distance difference between the two states, which can be calculated by the formula $\Delta d = \frac{2f \cdot |d_1 - d_2|}{d_1 + d_2}$, where $f$ represents The focal length of the camera, $d_1$ and $d_2$ are the depth values in the neutral state and the action unit state respectively.

We show the average amplitude changes of facial landmarks in Figure~\ref{fig:hotmap}. 
Spots with maximum movement amplitudes are concentrated in the active mouth area and regions influenced by skeletal muscles, which cannot be easily integrated into smart glasses though.
Within the area covered by the glasses, the intensity of facial muscle changes is most significant in the temporalis muscle above the zygomatic arch, which are marked with red circles in the figure.
Through facial anatomy analysis, we believe that there are two primary factors: first, this area serves as the junction between the upper and lower halves of the face.
When facial muscles contract, the corresponding skin stretches and retracts; 
second, the underlying temporalis muscle is a crucial component of the skeletal musculature.
Although it does not directly cause skin expansion and contraction, it induces subtle vertical skin movements.
In addition, we also counted the amplitude response of this position between different AUs. From Figure~\ref{fig:muscalebar}, we can see that AU26 has the largest range of motion, and AU5 has the smallest range of motion. But the amplitudes of all AUs are greater than 0.88, which fully demonstrates the potential of this location for sensor placement.

%In order to unify the measurement of these information, we need to convert the depth information from actual to pixel distance. Relying on the correlation between camera pixels and real-world distances, the pixel depth distance difference $\varDelta d_c$ can be calculated by the following formula:

%\begin{equation}
%\varDelta d_c = \frac{2 f |d_{neutral} - d_{AU}|}{d_{neutral} + d_{AU}}
%\end{equation}

%where $f$ is the focal length of the camera, $d_{neutral}$ and $d_{AU}$ represent the real-world depth in the neutral expression and AU state respectively.
%Here we ignore the difference between the camera's internal parameters $f_u$ and $f_v$ and unify them to $f$.
%The difference lies only in whether the depth mapping occurs in u-axis or v-axis direction in the image pixel coordinate system, with both remaining at the pixel level.

We then taped an IMU to this spot and observed the signal changes while a user made different expressions. 
The signal responses of the left and right measuring IMUs on AUGlasses are shown in the figure~\ref{fig:ex_imu}.
%We observed that during participants' expression of facial movements, the gyroscope responded mainly at the beginning and end phases of facial movements, which corresponded to instantaneous changes in angular velocity during the stretching and contraction of the muscles.
%Its peak value indicates the intensity of facial muscle contractions.
We noticed that the accelerometer accurately reflects the state of facial movement.
This is because the IMU's pose changes when the skin deforms, which leads to acceleration variations across three axes. 
%The gyroscope, on the other hand, mainly 
%Although the accelerometer was originally used to measure the acceleration of an object, due to the presence of gravitational acceleration, when the IMU changes position due to skin deformation, its direction of gravitational acceleration in the IMU coordinate system will also change, resulting in an acceleration variable.
%When participants maintain specific facial movements, the amount of change in this acceleration can represent the intensity of different facial movements, providing the possibility for us to continuously monitor facial movements.
To further verify this, we asked participants to express their facial expressions slowly.
In Figure~\ref{fig:browup}, the participant's expression slowly transitioned from neutral to eyebrow raising, and gradually returned to the neutral state from maximum intensity eyebrow raising.
The accelerometer values correspondingly increased and decreased with the facial movements.
This confirmed the feasibility of using acceleration data from IMU sensors for continuous tracking of facial movements.
% We did not use the gyroscope data to avoid excessive drifting, which can greatly impact our system's performance for long-time continuous tracking.  

\section{Implementation} \label{hardware_design}
% supporting structure design - hardware - software - power profile

With the sensor placement determined, we can design our smart glasses prototype. 
In this section, we first explain our design of the IMU supporting structure, which ensures tight yet comfortable contact of the IMU sensor to the skin around the temporal area.
Then we implement a moving artifact removal mechanism so that the IMU signal changes caused by head movement are filtered out.    

\subsection{IMU Supporting Structure Design and Manufacturing}
\begin{figure}[t]
  \centering
  \includegraphics[width=\linewidth]{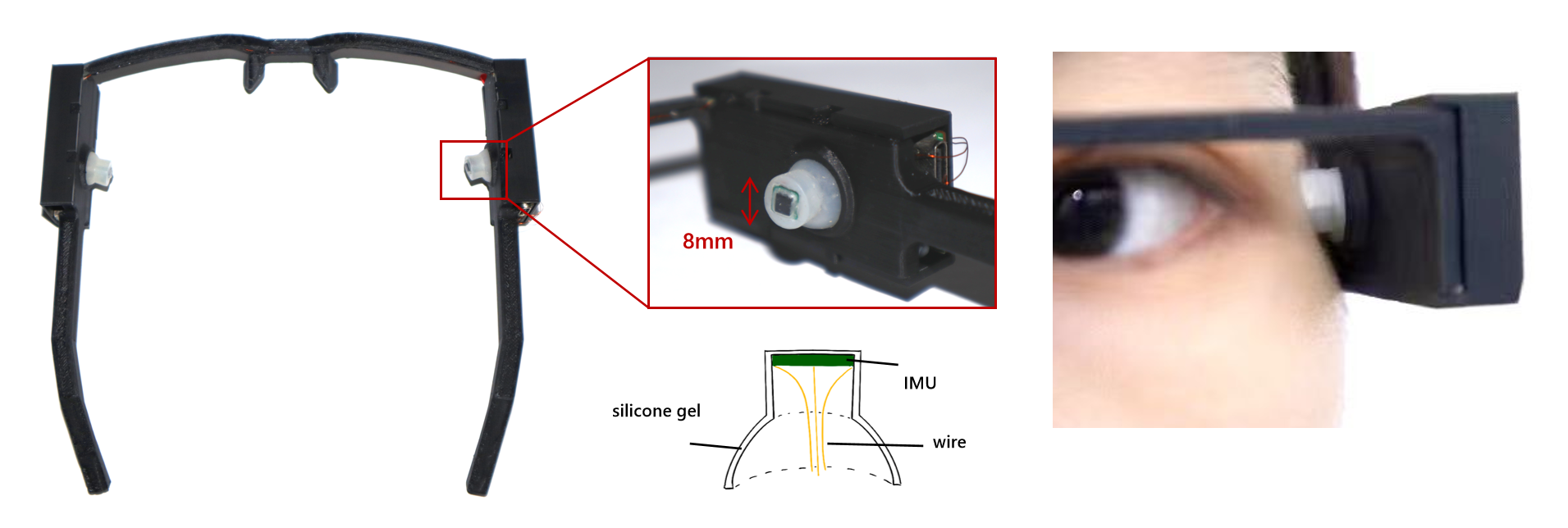}
  \caption{Prototype of AUGlasses.}
  \label{fig:Prototype}
\end{figure}

It is a challenging task to design a structure that can press the IMU tightly against the skin to maximize signal-to-noise radio, while flexible enough at the same time to avoid assert additional forces on the skin.
The structure also need to be easily integrated into smart glasses and small enough to avoid attracting attention. 
Inspired by the joystick structure on game controllers, we designed a supporting structure somewhat like a joystick controller on the leg of the smart glasses (Figure~\ref{fig:Prototype}).
We design a holster structure on the far end of the stick and the IMU chip is inserted into the gel to avoid direct contact with the skin. 
The rear features a bowl-shaped structure enabling telescopic and wide-angle sensor movement.

We use bio-compatible silicone gel Ecoflex-Gel2~\footnote{\url{https://www.smooth-on.com/products/ecoflex-gel-2/}} for safe and comfortable skin contact.
The gel is also adhesive to skin after cured, which ensures a stable fixed contact point on the skin.
We print internal and external molds of the structure using 3D printing with PLA material, then injecting the Ecoflex gel into the mold with a 1A:1B ratio. 
Thanks to its softness, the silicone gel not only supports the IMU sensor stably but also responds to minor expansions and contractions of facial muscles with high sensitivity.
\subsection{Smart Glasses Frame Design and Manufacturing}
The smart glasses frame has a compartment structure on each leg to hold the PCB board (43mm x 20mm) and the 350mAh lithium-polymer battery respectively.
The frame is manufactured by PLA 3D printing. 
To accommodate different face sizes and shapes, the frame has a curvature above the nose so that it can be slightly bent. 
We also apply Blue-Tack on the nose pads to ensure a comfortable wearing experience for different people.
Blue-Tuck is also used for counterweighting to balance the weight on both sides of the glasses.
The entire glasses weigh only 43 grams without the lens. 

The PCB board has two major components-a Bluetooth chip and an IMU chip. 
We use a Nordic nRF52840 module for data collection, assembly, and wireless communication. 
The module directly interfaces with three IMUs, extracting data using the SPI protocol.
Aside from the two IMUs collecting skin deformation data, one IMU is added on the PCB board for moving artifact removal purpose. 
The two off-board IMUs are connected with the PCB using 0.15mm enameled wire, discreetly concealed within the upper beam of the frame.
We use low-power BMI160 six-axis IMUs with a minimal size of 2.5 mm × 3.0 mm for all three IMUs. 
The PCB boards also include power management circuits and a Type-C port.

Data is transmitted in real time to a laptop.
To facilitate data processing in later experiments, we added a red LED to indicate the starting of data collection, which also enables convenient synchronization of IMU and video signals.  
To reliably capture the moment the red indicator light activates, our prototype first connects with the PC via Bluetooth, waits five seconds, then simultaneously turns on the red light and begins transmitting IMU data.
This procedure ensures the starting timestamp of the IMU data collection precisely aligns with the video's initial frame featuring the red light.

\begin{table}[h]
    \centering
    \caption{System power consumption of existing relevant sudies.}
    \label{tab:sensor_power}
    \begin{tabular}{|c|c|c|}
        \hline
        \textbf{Studies} & \textbf{Sensors} & \textbf{Power Consumption} \\ \hline
        SPIDERS\cite{nie_spiders_2020} & IR cameras, proximity sensor, IMU & 759.90 mW \\ \hline
        NeckFace\cite{chen_neckface_2021} & Cameras & 4W \\ \hline
        C-Face\cite{chen_c-face_2020} & Cameras & >4W \\ \hline
        BioFace-3D\cite{wu_bioface-3d_2021} & Biosensors & 138 mW \\ \hline
        EarIO\cite{li_eario_2022} & Acoustics & 153.7 mW \\ \hline
        \textbf{AUGlasses} & \textbf{IMUs} & \textbf{49.95mW} \\ \hline
    \end{tabular}
  
\end{table}

\subsection{Power Consumption}
To assess the feasibility of deployment into commercial smart glasses, we measured the power consumption characteristics of AUGlasses.
We used a multimeter to measure the battery's current draw and calculated the AUGlasses' power consumption using the formula: voltage x current.
Specifically, if the system is on standby, that is, no data is being transmitted via Bluetooth, the power consumption of AUGlasses is 35.15 mW (9.5 mA @ 3.70 V).
When the AUGlasses transmit 400Hz IMU data in real time, the power consumption of the entire system is 49.95mW (13.5mA @ 3.70 V).
This performance indicates that the system can deliver 25.9 hours of continuous data on a 350mAh lithium battery, satisfying the requirements of most applications.
% both of sensing power and communication power
We reviewed the power consumption of recent continuous face tracking studies and present the overall system power consumption (including sensing and communication power consumption) for each study in Table~\ref{tab:sensor_power}.
Our analysis shows that AUGlasses exhibits the lowest power consumption among all evaluated sensing technologies, and its energy consumption is 63.8\% lower than the next lowest Bioface-3D.

\section{Method}
% sensing pipeline - preprocessing - model structure - 3D reconstruction in Unity 
We show the overview of our entire system in Figure~\ref{fig:overview}.
First, there is a quick calibration process to ensure the IMUs can capture skin movement effectively.
Then the data goes through the pre-processing stage for normalization and moving artifact removal. 
The pre-processed IMU data is fed into a transformer-based deep learning model. 
The input of the model has two parts: 1). IMU data from the \textbf{current} frame; 2).the predicted AU intensities from the \textbf{last} frame.
In this way, the model estimates the AU intensities of the current frame with information both from the IMU sensor and from the internal relationship of consecutive frames, which improves the prediction accuracy. 
The predicted intensities of 14 AUs are then further processed for 3D facial reconstruction in Unity. 

%Due to the lack of existing datasets, we recruited 13 participants to collect facial motion-related IMU data and videos.
%Initially, we align the two data streams (see Section~\ref{Synchronization}) before feeding them into the system pipelines.
%Following video data preprocessing (see Section~\ref{Video_Processing}), we use the OpenFace tool to extract AU intensity labels for network training.
%IMU data stream preprocessing can be found in Section~\ref{IMU_Processing}.
%Subsequently, we use the RealTimeFormer model (see Section ~\ref{model}) to train on IMU data and AU intensity labels.
%In the testing phase (see Section ~\ref{real_time}), the IMU signal is fed into the trained network in real time, extracting AU intensity without visual input.
%Through the mapping between AU and blendShape, we use the blendShape corresponding to 14 AUs to achieve 3D facial reconstruction.
%In the evaluation section, we report the reconstruction results(see %Section~\ref{eva_Reconstruction}) in detail, demonstrating the excellent performance of AUGlasses in 3D reconstruction.

% \begin{figure}[t]

\subsection{Pre-processing} \label{preprocessing}

\subsubsection{Signal Normalization}

Due to the face shape and facial muscle differences, the IMU data for different AU strengths is different across users.   
To address this challenge, we ask users to collect personalized data for the first time using our system.
Among all AUs, we select Jaw Drop (AU26) as it is easy for users to understand and imitate.
The user is asked to express AU26 to the maximum extent and repeat three times.
Define the maximum intensity of the user's expression of AU26 on the i-th attempt as $au_i$. 
The maximum absolute signal values recorded by the accelerometer and gyroscope during this attempt are $acc_i$ and $gyro_i$, respectively.
The maximum values, $acc_{max}$ and $gyro_{max}$, of the user's accelerometer and gyroscope data are then calculated as follows:

\begin{equation}
{\Large acc/gyro}_{max} ^{left/right} = \sum_{i}^{3}(\frac{5acc_i}{au_i})
\end{equation}

Then, we normalize the signal using $(x_i-{min})/({max}-{min})$ so that the predicted AU values are standardized across users, which will facilitate the cross-user performance of our model. 

\begin{figure}[t]
  \centering
  \includegraphics[width=\linewidth]{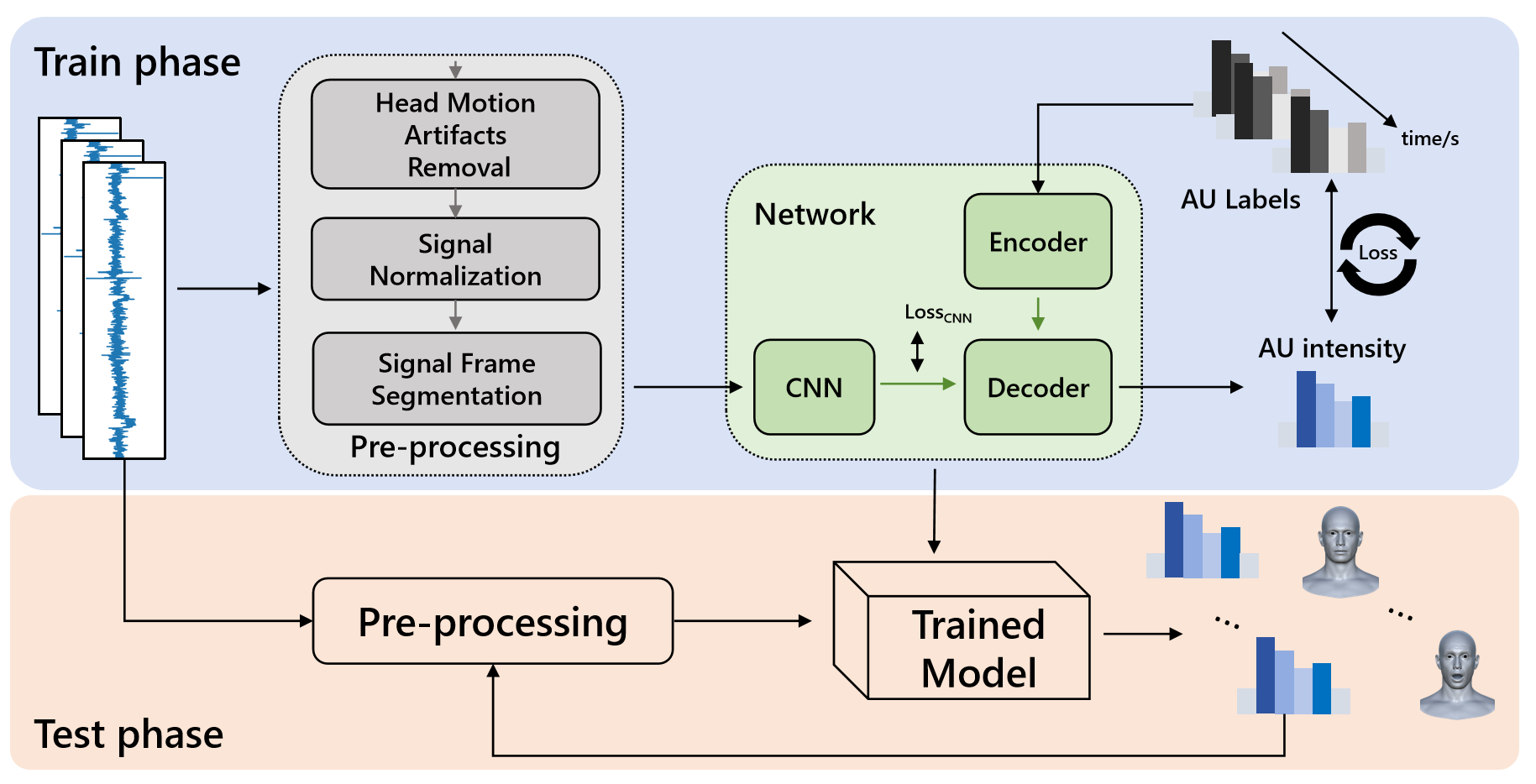}
  \caption{AUGlasses system overview.}
  \label{fig:overview}
\end{figure}

\subsubsection{Artifacts Removal} \label{Artifacts_rv}
\begin{figure}[t]
    \centering
    \captionsetup[subfigure]{justification=centering} % 设置子图的标题为居中
    \begin{subfigure}[b]{0.49\linewidth}
        \includegraphics[width=0.96\linewidth,height=4cm]{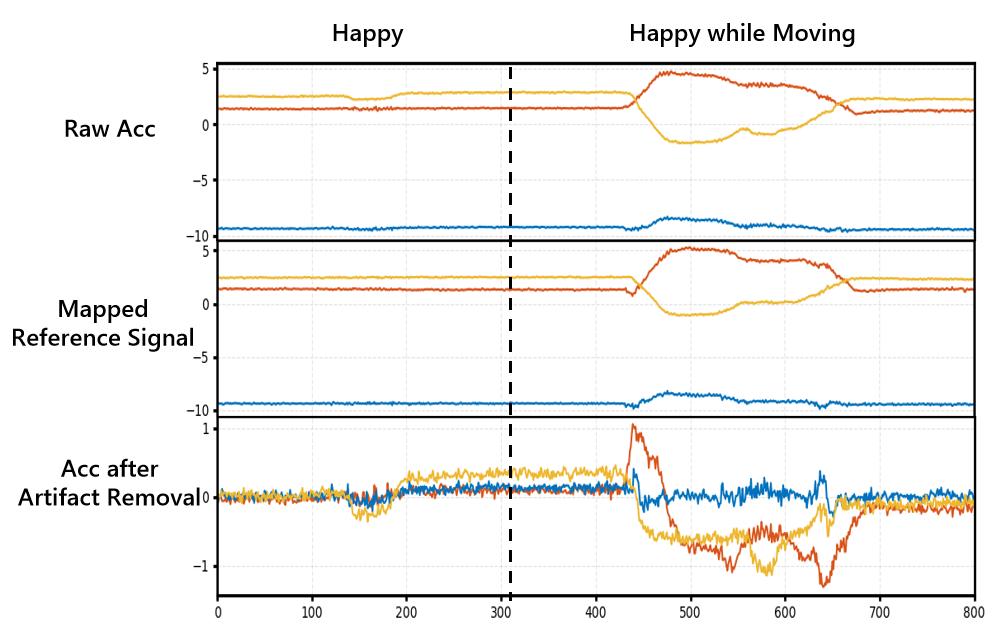}
        \caption{Accelerometer artifact removal.}
        \label{fig:acc_art}
    \end{subfigure}
    % \hfill % 在子图之间添加空格
    \begin{subfigure}[b]{0.49\linewidth}
        \includegraphics[width=0.96\linewidth,height=4cm]{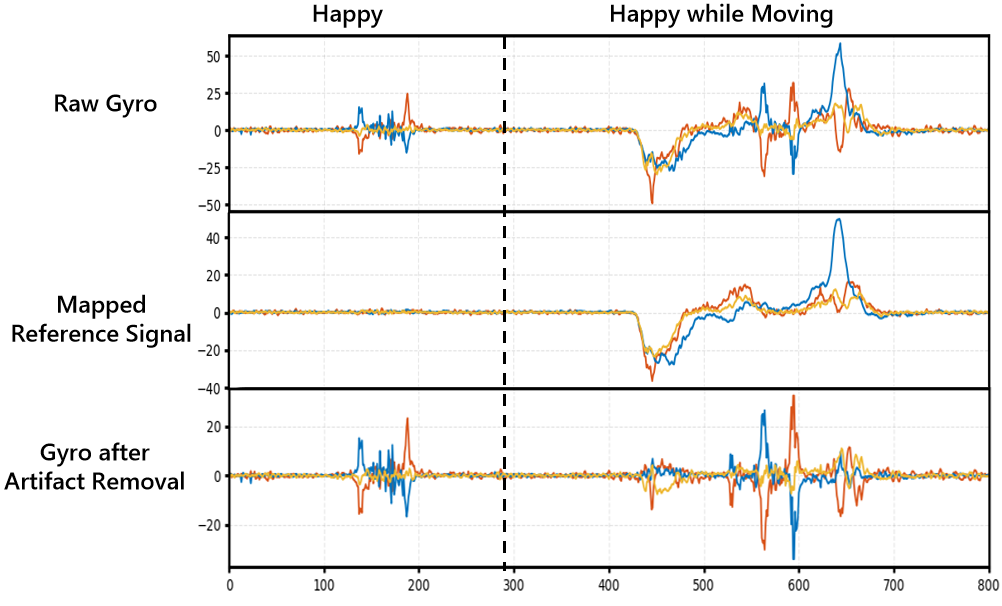}
        \caption{Gyroscope artifact removal.}
        \label{fig:gyro_art}
    \end{subfigure}
    \caption{Performance of artifact removal.}
    \label{fig:artifact}
\end{figure}

Our system has data from three IMUs, two in contact with facial skin and one on the PCB board in the glasses frame.
We first pass the signals through a low-pass filter set at 10Hz to eliminate high-frequency noise.
Given that the IMU is mounted on the head, it records both facial muscle and head movements.
To eliminate the significant interference caused by head movement, we use the affine transformation method for artifact elimination.
Specifically, given the six-axis signals ($ax$, $ay$, $az$, $gx$, $gy$, $gz$) of the onboard IMU, the transformed signals ($ax_t$, $ay_t$, $az_t$, $gx_t$, $gy_t$, $gz_t$) can be obtained by the following formula:
\[
\begin{bmatrix}
ax_t \\
ay_t \\
az_t \\
gx_t \\
gy_t \\
gz_t 
\end{bmatrix}
=
\textit{\textbf{R}}\textit{\textbf{x}} + \textit{\textbf{t}}
=
\begin{bmatrix}
r_{11} & r_{12} & r_{13} & r_{14} & r_{15} & r_{16} \\
r_{21} & r_{22} & r_{23} & r_{24} & r_{25} & r_{26} \\
r_{31} & r_{32} & r_{33} & r_{34} & r_{35} & r_{36} \\
r_{41} & r_{42} & r_{43} & r_{44} & r_{45} & r_{46} \\
r_{51} & r_{52} & r_{53} & r_{54} & r_{55} & r_{56} \\
r_{61} & r_{62} & r_{63} & r_{64} & r_{65} & r_{66}
\end{bmatrix}
\begin{bmatrix}
ax \\
ay \\
az \\
gx \\
gy \\
gz
\end{bmatrix}
+
\begin{bmatrix}
t_1 \\
t_2 \\
t_3 \\
t_4 \\
t_5 \\
t_6
\end{bmatrix}
\]
where $\textit{\textbf{R}}$ is the affine matrix and $\textit{\textbf{t}}$ is the translation matrix.
To determine $\textit{\textbf{R}}$ and $\textit{\textbf{t}}$, we capture a 1-second data segment and solve the corresponding equation.
We extract the initial segment of the signal for mapping purposes.
Such a mapping is carried out each time the head is detected to be still.
This is because the IMU data will drift and the mapping will be less effective as time goes by. 
To that end, we use the threshold method to continually update $\textit{\textbf{R}}$ and $\textit{\textbf{t}}$, thereby eliminating drift signals when the user remains stationary.
Specifically, we apply non-overlapping sliding window processing to the signals.
When all IMU signals are lower than the set threshold, we capture a 1-second data segment to remap the signals. 
Through this affine transformation, we effectively aligned the artifact signals of the measurement IMU and the onboard reference IMU, and then obtained pure facial muscle signals through differential operations.
Figure~\ref{fig:artifact} shows an example of our artifact remove method.  

% \subsection{Pre-processing} \label{IMU_Processing}
% \begin{figure}[t]
%   \centering
%   \includegraphics[width=\linewidth]{img/artifacts.jpg}
%   \caption{artifacts.}
%   \label{fig:artifact}
% \end{figure}

\subsubsection{Signal Segmentation}
Directly inputting continuous timing signals into the transformer architecture would result in an excessive number of model parameters.
Inspired by the Bioface\cite{wu_bioface-3d_2021} processing method, we divided the IMU signal into several short patches.
Each patch is paired with the AU labels of the corresponding frame, allowing us to derive the information for each AU label from the characteristics of the small IMU signal segment.
Specifically, we divide the IMU signal into overlapping short segments.
The length of each IMU signal segment is set to 200 data points, which is equivalent to 0.5 seconds.
%The frame length is set to 60 frames.

\subsection{Model Architecture and Training Strategy} \label{model}
In this study, we design an innovative architecture to predict AU intensity, illustrated in Figure~\ref{fig:structure}.
% \subsubsection{Yhy Is realTimeFormer?}
% Numerous studies employ the transformer architecture for time series data prediction.
% For instance, certain researchers\todo{} used EMG signals to predict blood pressure levels but limited their analysis to the prediction error of signal segments without venturing into real-time predictions.
% Using their architecture for real-time prediction, the overlapping sliding windows might yield inconsistent results for adjacent windows in the same period; the entire sliding windows could emerge gaps in predictions between them.
% To address this challenge, the realTimeFormer architecture we proposed effectively overcomes the above problems and achieves real-time analysis and prediction of continuous signal flows.

\subsubsection{CNN-based Feature Extraction}
We feed the IMU patches into a convolutional neural network (CNN) to extract features.
The CNN architecture includes four 1D convolutional layers, two max-pooling layers, and a fully connected layer.
To minimize over-fitting, we incorporate a dropout mechanism into the network.
In this architecture, the kernel size of each convolutional layer is set to 3 and stride is set to 1 to capture more information by continuously increasing the number of channels.
The kernel size of the pooling layer is set to 2, and the stride is 2.
The dropout layer's ratio is set to 0.5.
The network uses LeakyReLU as an activation function.
Each IMU segment, sized 200x12, is fed into this CNN architecture, with the final output dimension being 1x106.

\subsubsection{Transformer Structure}
We then conduct value embedding and position encoding on the time series data derived from the CNN output (size 60x106).
Unlike the conventional practice~\cite{ullah_optimized_2024} of directly inputting sensor data into the encoder layer, we innovatively feed this data into a six-layer transformer decoder structure.
It is important to note that the cross-attention mechanism requires information from the output of the encoder.
In our design, the first 15 AUs are considered historical information, and the last 45 AUs serve as the target length for predictions.
In the encoder branch, we use historical AU label information as input.
To synchronize the model with the timing of the IMU signal and AU, we apply the strategy in PatchTST~\cite{nie_time_2023} to extend the encoder input to a total length of 60 using zero padding.
After value embedding and position encoding, the processed AU time series segments are fed into a six-layer encoder structure, then transferred to the decoder to extract cross-attention information.
At the output of the decoder, we remap the data back to 14 AUs' channels through a linear layer to finalize the prediction.

Regarding model parameters, we set $\text{d\_model}$ to 256, $\text{n\_head}$ to 8, $\text{d\_ff}$ to 512, and the dropout ratio to 0.1 in both the encoder and decoder.

\begin{figure}[t]
  \centering
  \includegraphics[width=\linewidth]{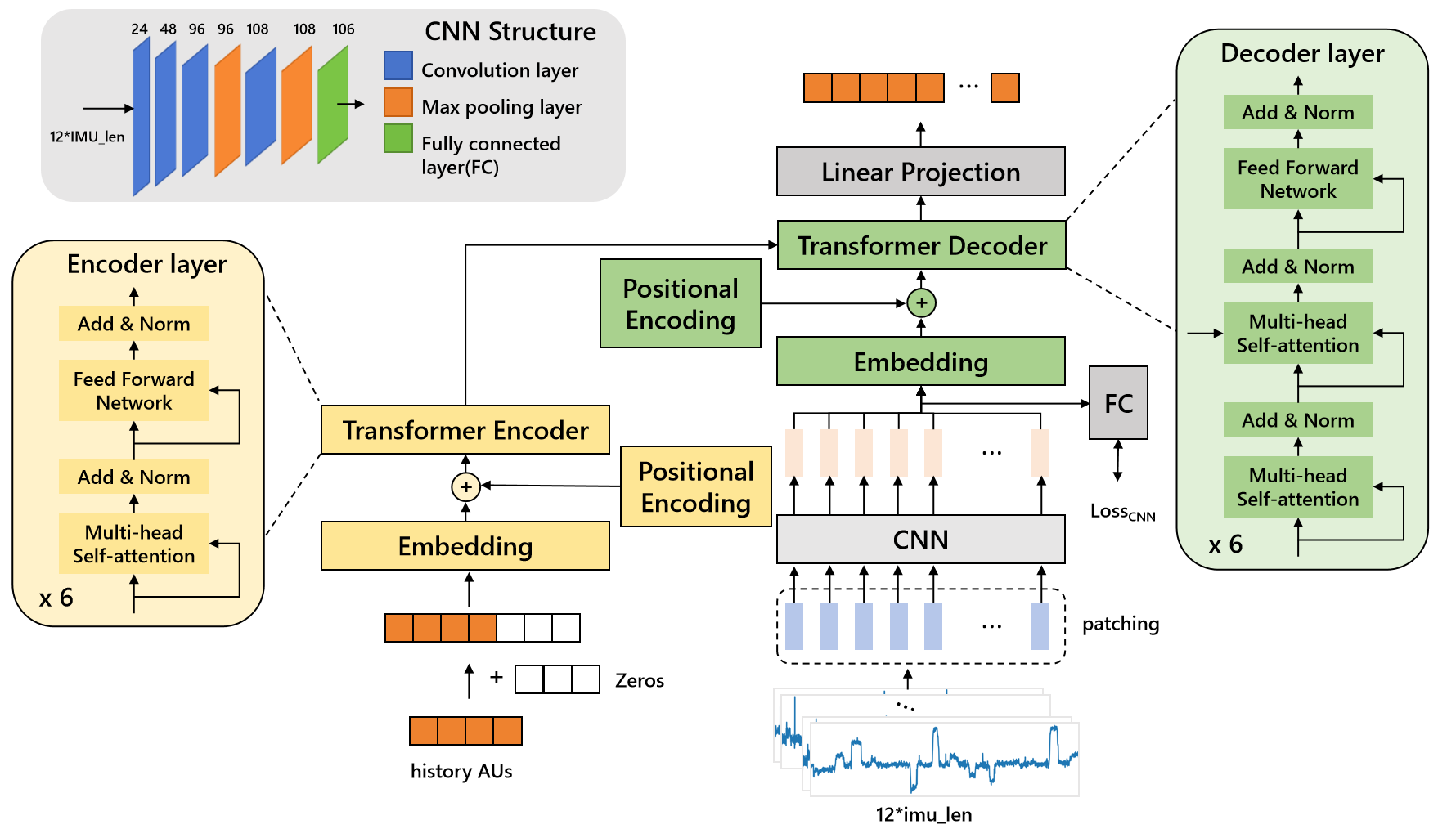}
  \caption{Illustration of model structure.}
  \label{fig:structure}
\end{figure}

\subsubsection{Training Strategy}
In traditional Transformer models, training with teacher forcing introduces exposure bias, causing the model to learn from the actual output instead of its predictions.
This method leads to rapid convergence but may result in poor inference performance due to accumulated errors.
Several methods, such as Scheduled Sampling~\cite{liu_confidence-aware_2021} and beam search~\cite{leblond_machine_2021}, are employed to mitigate this issue.
However, Unlike other applications with relatively long output intervals, facial reconstruction demands more frequent outputs, exacerbating the impact of exposure bias.
To address this challenge, we propose "prefix conditional sequence forecasting," which conditions the model on a prefixed sequence of observed data rather than relying on the immediate past ground truth.
This approach overcomes the limitations of standard autoregressive models by enabling broader contextual incorporation, reducing exposure bias, and enhancing predictive accuracy for tasks with rapid and subtle changes.
% [把你的方法写在这里]

% 要点：1: prefix 指的是我们的history长度固定。（15） 2. segment level指的是，我们预测的时候，在45层面上，全部预测，不进行自回归。

\subsection{Training Setup and Settings}
The model training process is shown in the top part of Figure~\ref{fig:overview}.
We train our model with AU ground truth calculated by OpenFace~\cite{baltrusaitis_openface_2016} from video data. 
OpenFace is a neural network-based face recognition toolkit that recognizes 17 AU intensities (ranging from minimal to maximal on a 5-point scale).
The video and IMU data are synchronized with the red LED light on the glasses.
The LED lights up to signify the start of the IMU data sequence. 
The first frame where the red pixels are identified is set to be the starting frame. 
The captured video is resampled for a uniform frame rate of 30 frames per second.

To enhance the performance of the CNN feature extraction module, we implement a multi-loss function strategy.
The entire network is trained through minimization of the following cost function:
\begin{equation}
L = E_{cnn} + E_{Former}
\end{equation}
where $\lambda_1$ and $\lambda_2$ are hyper-parameters, representing the weights assigned to each cost component.
$E_{Former}$ represents the loss of the entire network architecture, while $E_{cnn}$ denotes the CNN feature extraction module's loss.
The calculation involves flattening the CNN's output data and passing it through a fully connected layer to compute the mean square error (MSE) with the labels.
The overall network architecture's loss is similarly computed using MSE.
Through grid search, we found the optimal values of these hyper-parameters: $\lambda_1$=0.5 and $\lambda_2$=1.5.

In addition, the network is trained using the SGD optimizer, the learning rate is set to 0.001, the batch size is 32, and the learning rate scheduler is OneCycleLR.

\subsection{Real-time Estimation} \label{real_time}
In the training phase, we use ground truth AU values as input; in the initial phase of real-time reasoning, we use a threshold algorithm to first determine the static state of the user, and then replace the historical information with all zero values and input it to the encoder.
This is primarily to reset the accumulated prediction error.
Given the network's prediction interval is very brief, the system can swiftly capture the initial static state and commence real-time prediction unnoticeably to the user.

We use overlapping sliding window to intercept IMU signal segments in real time.
Assuming that the prediction time period is $N$ and the length of historical information is $s$, the window step size is set to $N-s$.
This configuration incorporates IMU data from the previous window into subsequent prediction windows.
Simultaneously, the prediction result spanning s duration at the current window's end is input as the next window's historical information.
% ####
% As shown in the figure xx,\todo{} realTimeFormer enhances signal continuity awareness and ensures prediction result consistency.
When processing the overlapping parts of the prediction windows, we replace the previous results with subsequent prediction results as real-time prediction output for further downstream task analysis.

\subsection{3D reconstruction}
In previous studies, researchers typically achieved 3D facial reconstruction by estimating key facial points\cite{martyniuk_dad-3dheads_2022} or blendshapes\cite{li_eario_2022}.
%AUGlases, due to its design focus on low power consumption and high concealment, the only two inertial measurement units (IMU) integrated in the device are not enough to support such a complex facial data capture task.
%Therefore, we turned to a new strategy, which is to use AU to represent facial motion, and explore the possibility of three-dimensional facial reconstruction based on AU data.
%AUs and blendshapes represent two distinct methods of expressing facial movements.
%By understanding their definitions, we can identify the related aspects between them for mapping purposes.
%Using examples from relevant websites\footnote{https://melindaozel.com/arkit-to-facs-cheat-sheet/}, we individually map 14 AUs to their corresponding blendshapes.
%Due to the absence of a direct match for Lips Part(AU25), we exclude it from facial AU reconstruction.
%Fortunately, even without AU25, facial AU reconstruction can still be effectively performed through the relevant AUs provided by openface such as Lip Corner Puller(AU12), Jaw Drop(AU26), etc.
We render a virtual face in Unity based on the estimated AU intensities. 
To be compatible with existing Unity plugins, we convert the predicted AU intensity sequence into blendshape parameters in real time and import it into Unity for animation processing in real time.
Since the blendshape threshold in Unity is set to 0-100, we linearly scale the predicted intensity (original range 0-5) to the 0-100 scale for expression.
The face of an avatar is then reconstructed. 
%Next, we assemble the video frames in Unity, restore them to the original frame rate, and conduct a comparative analysis with the original video.

\section{Evaluation}
We conducted a lab-controlled user study to collect training data and evaluate our system's performance on facial reconstruction. 
With the trained model, we also carried out a series of micro-benchmark tests to further understand our system's performance during walking and for long-time continuous forecasting.  

\subsection{Experimental Design and Procedure}
\subsubsection{Experimental Settings} \label{ex_setting}

To evaluate the device's performance, we recruited 13 subjects (8 females, 5 males, ages 22 to 28) from the local institution.
Participants were informed of the video recording and signed the consent form, and each participant was awarded US\$13.80 for completing the experiment.
The experimental equipment includes AUGlasses prototype, HP ZBook Power G9 laptop, nrf52840 development board connected to the laptop via USB for bluetooth communication.
Initially, participants turned on the glasses and waited for a steady blue light to indicate Bluetooth connection to the computer.
When the red light turns on, it means the device starts collecting IMU data. 
At the same time, we used the built-in camera on the computer (720p resolution and 30fps) to capture video.

Before the experiment, we showed participants pictures of 7 expressions (sadness, anger, happiness, surprise, fear, disgust, and neutral) and 11 AUs so that they can successfully imitate them. 
Instead of just asking participants to perform all 14 AUs, we selected 11 AUs that most participants can perform accurately, and ask them to make 7 typical expressions (happy, sad, fear,disgust, anger, surprise, contempt) to collect multiple AU strengths simultaneously, which covers the 14 target AUs. 
This is because it is difficult for participants to perform certain AUs.
By collecting compound AU from expressions, we also reduced the study duration. 

Each participant completed two sessions of experiment. 
For each session, the participant went through three phases: calibration, individual AUs, and compound AUs (expressions).  
In the calibration phase, participants performed two actions — "Still" and "Jaw Drop (AU26)" — repeating three times.
They are reminded the start and end of actions, as well as rest periods through audio. 
Participants were instructed not to blink during expressions but could blink during rest periods.
In the individual AUs phase, participants performed the 11 AUs 10 times each in a random order, which took around 15 minutes. 
In the compound AUs phase, they expressed the 7 facial expressions 10 times each in a random order, which took about 7 minutes.
After completing all three phases, subjects removed the equipment and rested for two minutes before wearing it again to start the next session.

\subsubsection{Evaluation Metrics}
We report cross-session result for each user to understand the practical performance of a highly user-personalized model.  
We also report cross-user result to demonstrate our model's performance in generalizability across sessions and users. 
Metric-wise, we primarily use mean absolute error (MAE) as our evaluation metric.
Specifically, we calculated the MAE between the model's predictions for 14 facial AUs intensity and those predicted by OpenFace.
However, it is important to note that if OpenFace performs poorly in some specific cases, this may cause the error to be amplified.

The 3D face reconstruction performance is evaluated primarily by two metrics~\cite{wu_bioface-3d_2021}:

1) Mean Absolute Error (MAE): the absolute error between the reconstructed face and groundtruth face landmarks: $MAE(P,\hat{P}) = \frac{d_r}{d_f} \times ||p_i - \hat{p}_i||_2$, where $P$ and $\hat{P}$ are the ground truth and predicted landmark coordinates for each image, $d_f$ is the distance between the two outer canthus in the frame,
$d_r$ is the distance between the two outer canthus of the actual participant.

2) Normalized Mean Error (NME): the average error between reconstructed and groundtruth landmarks coordinates, normalized by the inter-ocular distance, represented by $NME(P,\hat{P}) = \frac{||p_i - \hat{p}_i||_2}{d}$, where $p_i$ is the $i$-th ground truth landmark coordinate in $P$; $\hat{p}_i$ is the $i$-th predicted landmark coordinate in $\hat{P}$, $d$ is the normalization factor.

\begin{figure}[t]
    \centering
    \captionsetup[subfigure]{justification=centering} % 设置子图的标题为居中
    \begin{subfigure}[b]{0.325\linewidth}
        \includegraphics[width=0.9\linewidth]{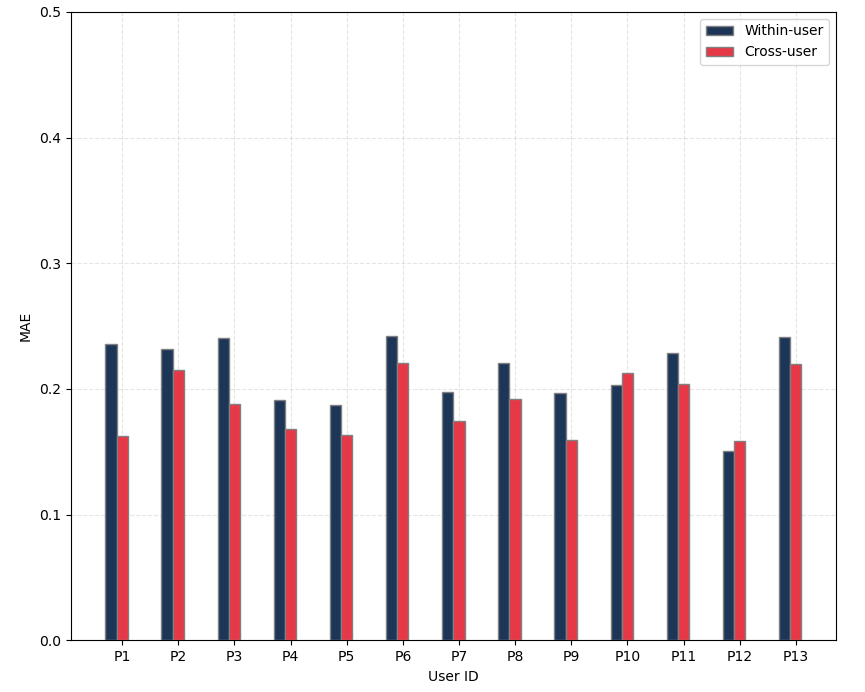}
        \caption{Per-participant AU tracking error.}
        \label{fig:aumae}
    \end{subfigure}
    \hfill % 在子图之间添加空格
    \begin{subfigure}[b]{0.325\linewidth}
        \includegraphics[width=0.9\linewidth,height=4cm]{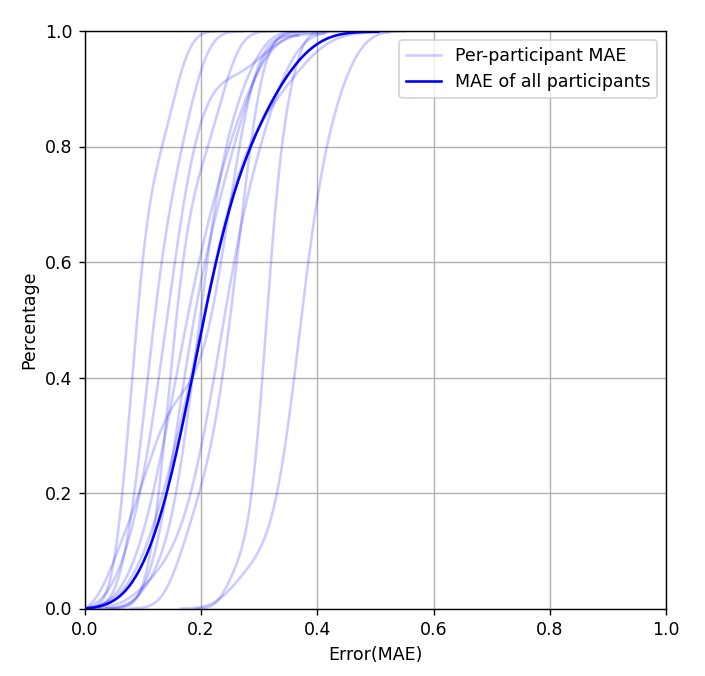}
        \caption{CDF of within-user setting.}
        \label{fig:CDFwithinuser}
    \end{subfigure}
    \hfill % 在子图之间添加空格
    \begin{subfigure}[b]{0.325\linewidth}
        \includegraphics[width=0.9\linewidth,height=4cm]{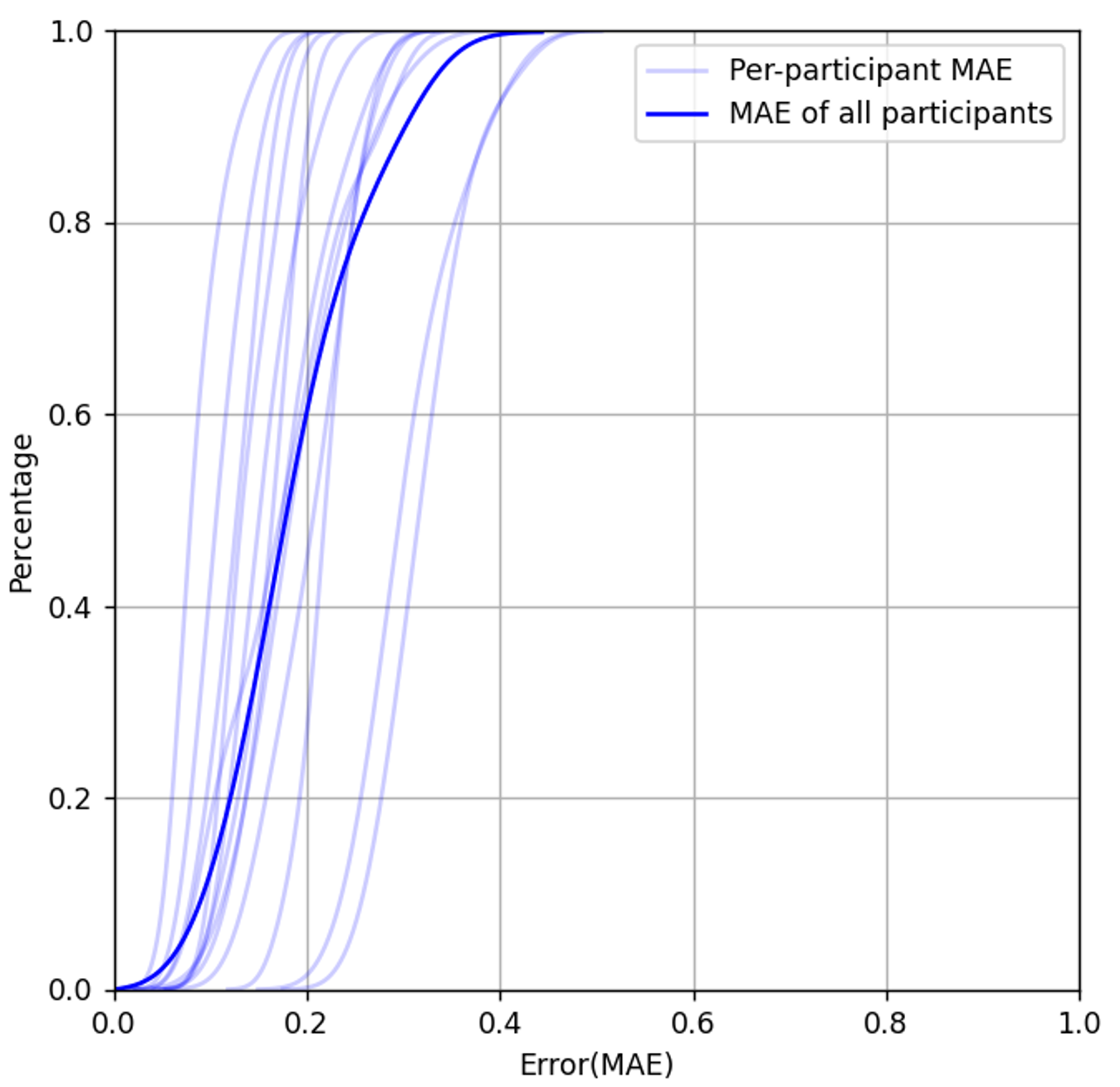}
        \caption{CDF of cross-user setting.}
        \label{fig:CDFCross-user}
    \end{subfigure}

    \caption{Performance of continuous facial unit tracking.}
    \label{fig:maecdf}
\end{figure}

\subsection{AU Estimation Results} \label{base_result}

We initially assess AUGlasses' AU intensity estimation across two model settings.
We train using all collected expressions and AU datasets under corresponding settings.
Figure~\ref{fig:maecdf} presents the average MAE and cumulative density function (CDF) for 14 AU intensity estimations from 13 participants in within-user (cross-session) and cross-user settings, demonstrating AUGlasses' effective detection of fundamental facial AUs.

\subsubsection{Within-user (Cross-session) Evaluation}
From the two sessions of data for each user, we randomly picked the training session and test the trained model on the remaining session.
Results~\ref{fig:aumae} show that AUGlasses achieves an average MAE of 0.21 (STD=0.028).
Among all participants, P12 had the best performance results with a MAE of only 0.15, while P13 had the highest at 0.24.
As can be seen from Figure~\ref{fig:CDFwithinuser}, 65.6\% of the AU predictions have MAE less than 0.25, showing a relatively high prediction accuracy.
%These results validate AUGlasses' ability to track basic facial movements.

\subsubsection{Cross-user Evaluation} \label{cross_user}
In the cross-user setting, we utilize a leave-one-out approach, employing one individual's data as the test set and data from the other users for training and validation.
We designate the first 80\% of the training participants' data for training purposes and the final 20\% for validation. 
Figure~\ref{fig:aumae} shows that the MAE of 13 participants is 0.19 (STD=0.025), the maximum MAE value is 0.22, and the minimum MAE value is 0.16.
Additionally, Figure~\ref{fig:CDFCross-user} indicates that 74.6\% of AUs achieved an MAE below 0.25.
Surprisingly, the cross-user performance is better than that of the within-user (cross-session) results.
There are several potential reasons for this.
First, the amount of training data from cross-user model is much greater than that of the within-user model.
A relatively large amount of training data is necessary to improve the performance of a transformer based model. 
Second, with only two sessions of data collected, it is highly possible that the within-user model over-fits on the training session, leading to inferior performance on the test session.
Third, the cross-user training data contains rich information with different contact points in different sessions of different users. Our model effectively learnt the shared features of such different conditions and discards those session-specific knowledge, leading to a superior performance of cross-user models. 
Such a phenomena clearly shows that our system has a low requirement on the placement precision of the sensor on the face, and can be readily generalized to more users. 

\begin{figure}[t]
    \centering
    \captionsetup[subfigure]{justification=centering} % 设置子图的标题为居中
    \begin{subfigure}[b]{0.325\linewidth}
        \includegraphics[width=0.9\linewidth]{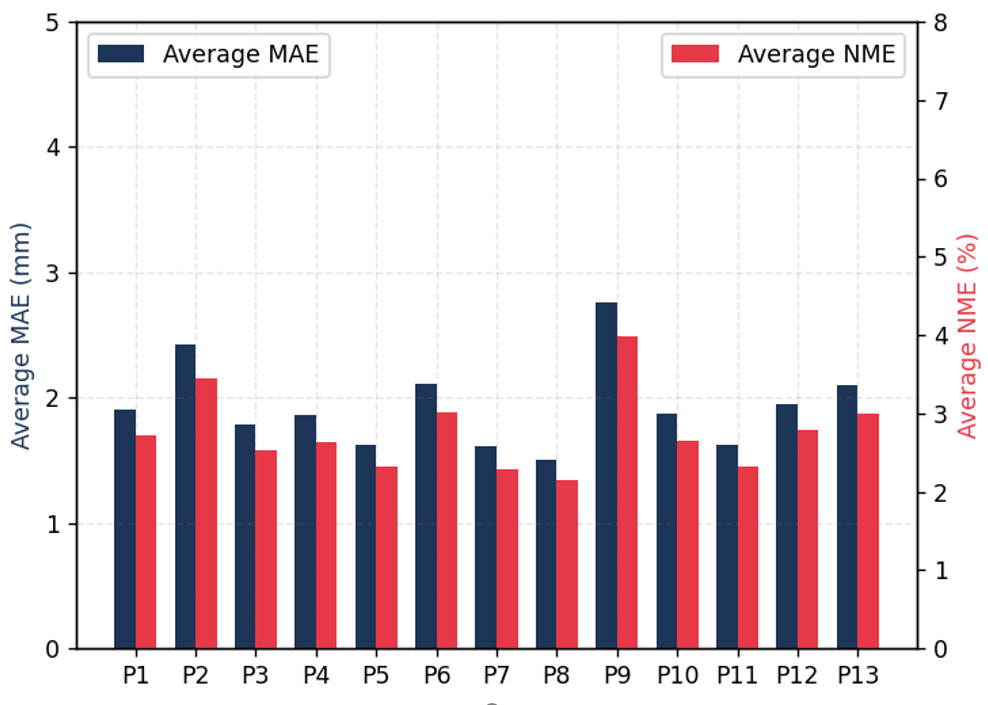}
        \caption{Per-participant landmarks tracking error(cross-user).}
        \label{fig:Per-participant_3d}
    \end{subfigure}
    % \hfill % 在子图之间添加空格
    \begin{subfigure}[b]{0.325\linewidth}
        \includegraphics[width=0.9\linewidth]{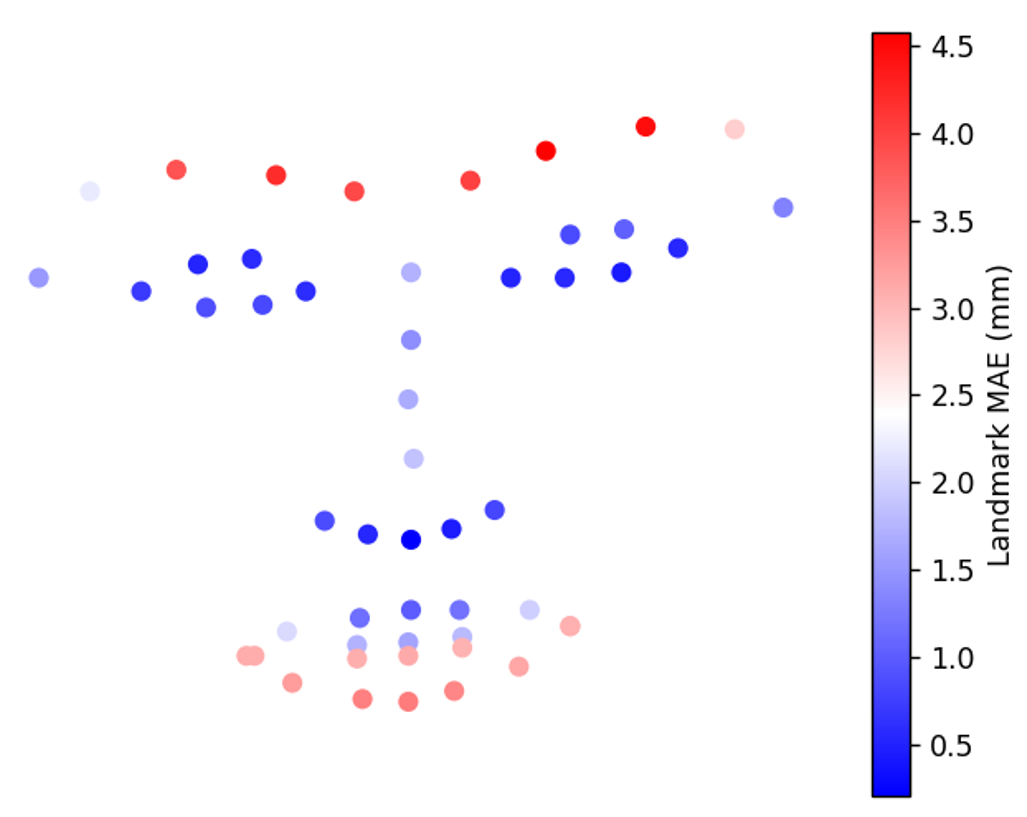}
        \caption{Visualization of average MAE(cross-user).}
        \label{fig:scatter_point}
    \end{subfigure}
    \begin{subfigure}[b]{0.325\linewidth}
        \includegraphics[width=0.9\linewidth]{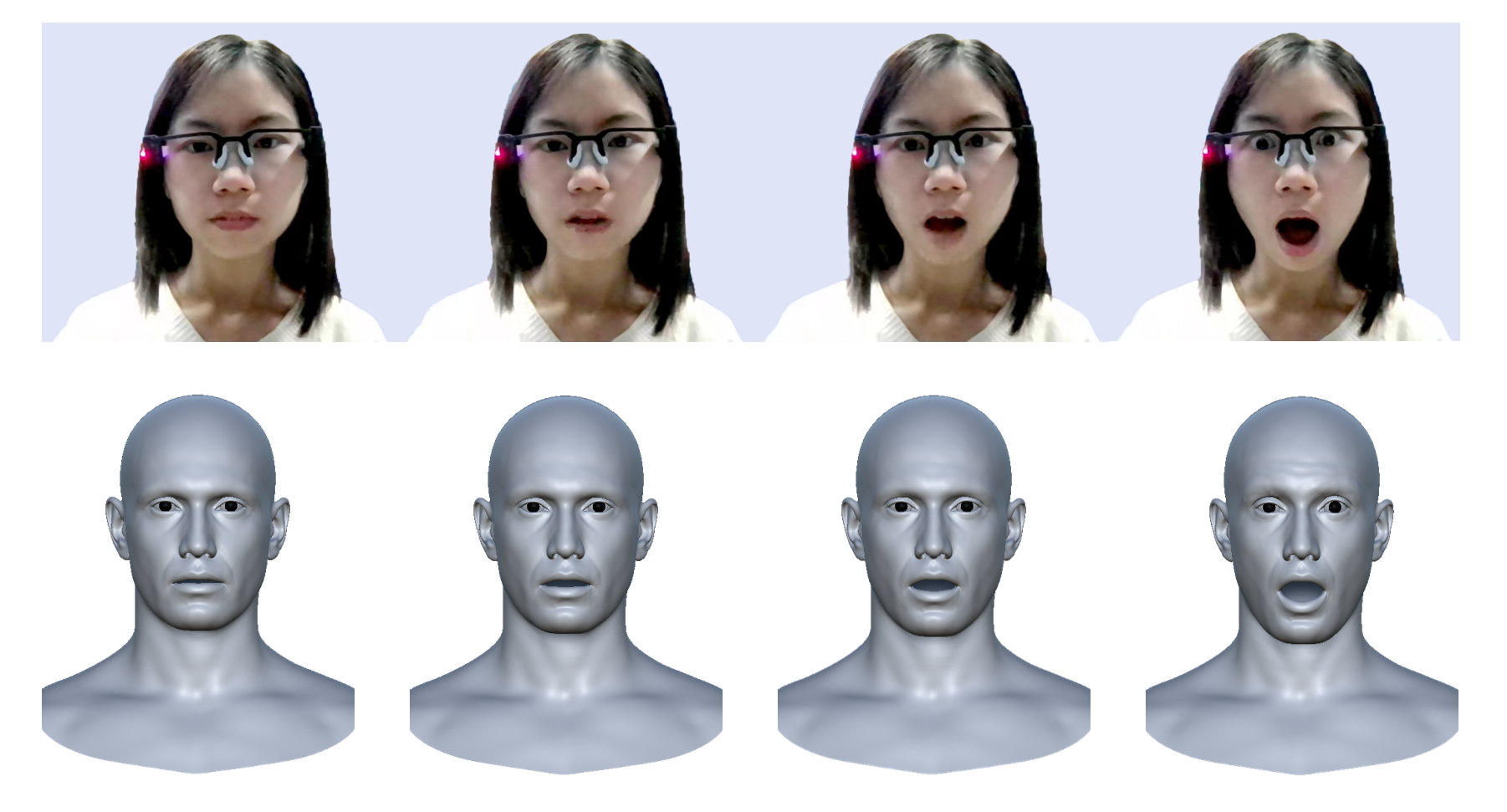}
        \caption{Example of continuous facial reconstruction.}
        \label{fig:Continuous_face_recon}
    \end{subfigure}
    \caption{Performance of 3D reconstruction.}
    \label{fig:3dconstruction}
\end{figure}

\subsection{3D Facial Reconstruction} \label{eva_Reconstruction}
\subsubsection{Facial Landmark Tracking}
%We utilized the expression and AU data from three consecutive rounds for a 51-point facial landmark evaluation.
Figure~\ref{fig:Per-participant_3d} displays the average MAE and NME for all 51 facial landmarks across with-user and cross-user settings for each participant.
Specifically, AUGlasses achieved an average MAE of 1.89 mm (std = 0.317) and an NME of 2.69\% (std = 0.0046) in within-user setting; the averages were 1.93 mm (std = 0.353) for MAE and 2.75\% (std = 0.0051) for NME in cross-user setting.
These results confirm the effectiveness of AUGlasses in reconstructing 3D faces based on AU data.

To demonstrate the error distribution, we visualized the average MAE for 51 facial landmarks under the more practical cross-user setting, as depicted in Figure~\ref{fig:scatter_point}.
The results show that the reconstruction errors for eyebrows and mouth are higher than for eyes (MAE = 0.70 mm) and nose (MAE = 1.06 mm), at 3.29 mm and 2.54 mm respectively.
This is mainly because the eyebrows and mouth are one of the most important expression parts of our common facial movements, and their movement variability is relatively high.
Despite this, their MAE values are still low enough for facial reconstruction.

To compare AUGlasses' performance with other studies, Table~\ref{table:NME} outlines the NME comparisons between AUGlasses and several recent facial landmark tracking technologies.
Among them, SAN  adopts a vision-based method and mainly utilizes public image datasets (300-W) to track 68 facial key points.
mm3Dface and BioFace-3D perform facial tracking based on millimeter wave signals and EMG signals respectively.
The comparative analysis reveals that AUGlasses performs comparably to existing vision-based solutions and excels in other sensing technologies.
% ####

\begin{table}[h!]
\centering
\caption{Comparison with existing solutions}
\label{table:NME}
\begin{tabular}{|c|c|c|c|c|}
\hline
\textbf{Methods} & \textbf{Dataset} & \textbf{Normalization} & \textbf{Landmarks} & \textbf{NME} \\
\hline
FIFA\cite{kar_fiducial_2024} & 300-W & Inter-ocular & 68 & 2.89 \\ \hline
3DDE\cite{valle_face_2019} & 300-W & Inter-ocular & 68 & 3.13 \\ \hline
SAN\cite{dong_style_2018} & 300-W & Inter-ocular & 68 & 3.98 \\ \hline
BioFace-3D\cite{wu_bioface-3d_2021}& Self-collected & Inter-pupil & 53 & 3.38 \\ \hline
mm3DFace\cite{xie_mm3dface_2023} & Self-collected & Inter-pupil & 68 & 3.94 \\ \hline
\textbf{AUGlasses} & \textbf{Self-collected} & \textbf{Inter-ocular} & \textbf{51} & \textbf{2.75} \\ 
\hline
\end{tabular}

\end{table}

\subsubsection{Continuous 3D Facial Reconstruction Results}
Figure~\ref{fig:Continuous_face_recon} shows an example of AUGlasses continuously reconstructing 3D facial movement. 
Through continuous 3D rendering of facial frames, we can observe that AUGlasses can smoothly perform continuous three-dimensional reconstruction and clearly show different degrees of facial changes. 
This confirms AUGlasses' exceptional proficiency in dynamically reconstructing facial movements.

\subsection{Subjective Results} \label{subjectiveresult}
\begin{figure}[t]
  \centering
  \includegraphics[width=\linewidth]{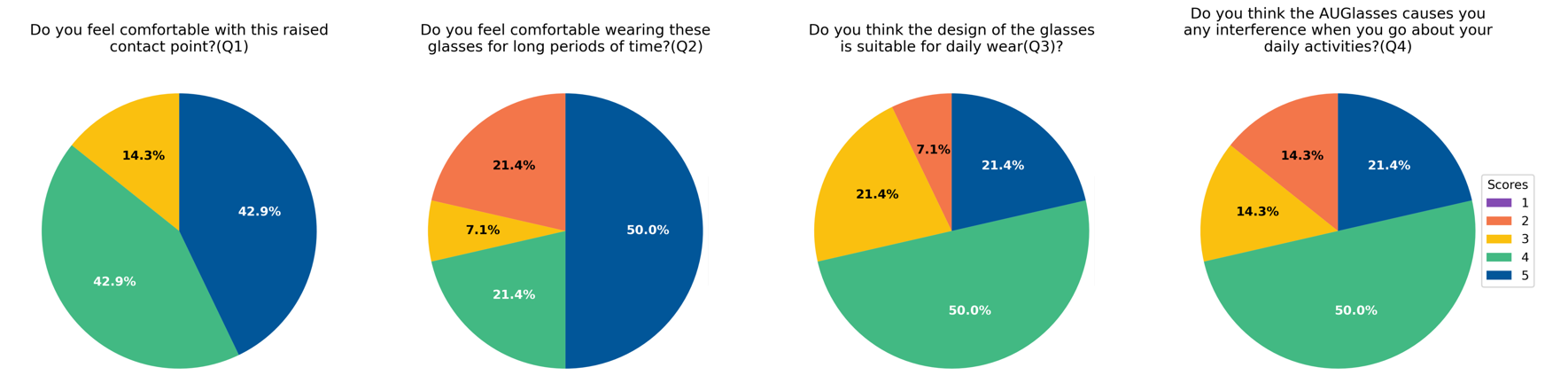}
  \caption{User experience questionnaire scores on AUGlasses.}
  \label{fig:pie}
\end{figure}

To assess the system's usability and comfort, participants completed a questionnaire after the experiments, as shown in Figure~\ref{fig:pie}.
Participants rated each question on a 5-point Likert scale, where 1 means “strongly disagree” and 5 means “strongly agree”.
Overall, the majority of participants had a positive attitude toward the AUGlasses.
85.8\% of the participants can barely feel the contact (Q1: MEDIAN=4, STD=0.73).
71.4\% of the participants felt that wearing it for a longer time than the experiment duration would be comfortable (Q2: MEDIAN=4.5, STD=1.24).
One participant commented the use of soft rubber for the protrusions, noting the contact points were almost imperceptible.
One participant did pointed out that the temples of glasses exerted pressure on the ears, which can be allieviated with more flexible frame materials in future design. 
71.4\% of participants felt that the design of the glasses was suitable for daily wear (Q3: MEDIAN=4, STD=0.86) and would not cause interference (Q4: MEDIAN=4, STD=0.97).
One participant commented,``Because two detection points are on the temporal part, during the actual wearing process, if you don’t pay attention deliberately, you will not realize that there are these contact points on both sides. It is relatively secretive.''
Participants also preferred the IMU-based sensing solution to visual or audio based solutions due to the perceived lower privacy risk. 
%The participant who rated it as 2 thought that wearing glasses was bothersome inconvenient, and she would not wear it when going out.
%This is attributed to the prototype's design, which is currently focused solely on functionality verification.
%To maintain the protrusion' elasticity, they are semi-squeezed when the participants' face is neutral, requiring participants to manually adjust them before wearing the glasses in the experiments.
%In the future, we can further optimize the overall structure and add micro springs and proximity sensors to the temples.
%The protrusion automatically shrink when the glasses are removed and expand to make close contact with the skin upon wearing.
%Furthermore, 64.3\% of participants preferred AUGlasses to sound detection methods (Q5: MEDIAN=4, STD=0.99).
%This preference is mainly due to perceived privacy risks with voice activation and the lightweight, compact design of the AUGlasses.
%This preference stems primarily from the perceived higher risk of privacy leakage with voice-activated devices. Additionally, users find the AUGlasses to be lightweight and compact.
\begin{figure}[t]
    \centering
    \captionsetup[subfigure]{justification=centering} % 设置子图的标题为居中
    \begin{subfigure}[t]{0.325\linewidth}
        \includegraphics[width=0.9\linewidth]{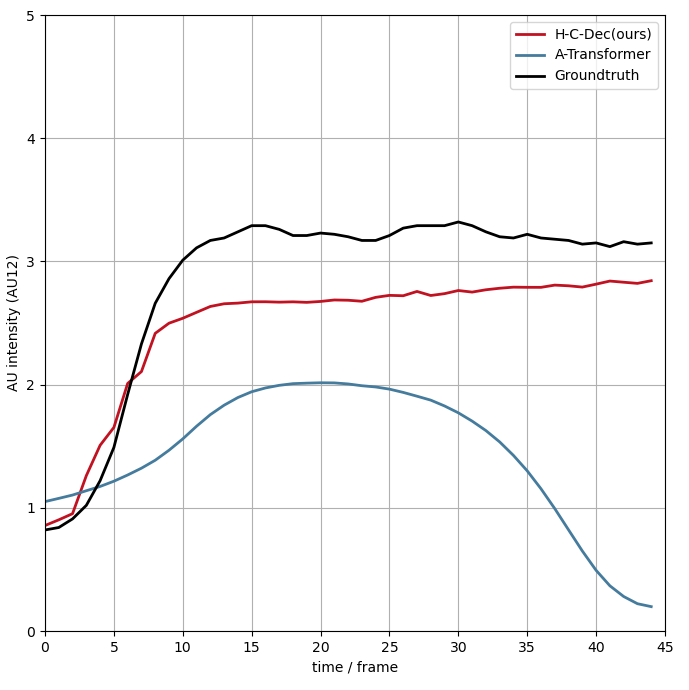}
        \caption{Forecast comparison A-Transformer vs. AUGlasses' model.}
        \label{fig:singleimu}
    \end{subfigure}
    % % \hfill % 在子图之间添加空格
    \begin{subfigure}[t]{0.325\linewidth}
        \includegraphics[width=0.9\linewidth]{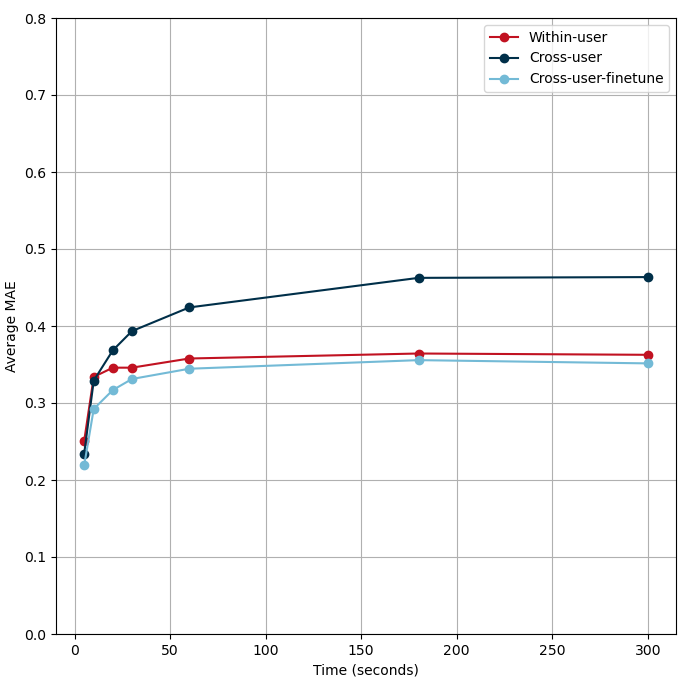}
        \caption{Performance of predictions over time.}
        \label{fig:model_dif}
    \end{subfigure}

    \caption{The performance of different experiments.}
    \label{fig:micro_benchmark}
\end{figure}

\subsection{Ablation Studies}
% Due to time cost, the following results are within-user results
To verify our model's effectiveness, we compared it with the typical autoregressive transformer method and conducted ablation studies to validate our model's superiority.
We run ablation studies using within-user settings and the results are presented in Table~\ref{table:Ablation}.

\textbf{History AU+C-Dec (H-C-Dec)} This serves as the backbone model for AUGlasses, depicted in Figure~\ref{fig:structure}. Unlike autoregressive models, this network incorporates zero-padding of historical AU data into the encoder and feeds IMU time series through a CNN layer to the decoder for prediction.

\textbf{Autoregressive-Transformer (A-Transformer)} The IMU signal is input into the encoder architecture through the CNN layer; the decoder's self-attention layer masks and predicts the AU time step after the current position.
It should be noted that the CNN, encoder, and decoder architectures are identical to those in the AUGlasses backbone, but the A-Transformer is trained autoregressively.
Results indicate that the A-Transformer's average AU MAE is 0.267.
Additionally, we present a comparison of the prediction curves between A-Transformer and H-C-Dec in Figure~\ref{fig:model_dif}.
This shows that for timing prediction tasks with high refresh rates, A-Transformer produces serious exposure bias in a short period of time.

\textbf{CNN+encoder (C-Enc)} To assess the historical information component's effectiveness, we remove the network's cross-self-attention layers, eliminating  the branches of historical information.
The table reveals that C-Enc records an average AU MAE of 0.363, which is 0.096 lower than that of the A-Transformer, underscoring the importance of historical information in enhancing prediction accuracy.

\textbf{C-Enc + History AU (C-Enc-H)} A typical method in time series prediction involves inputting the sensing signal into the encoder and using predicted historical information as the decoder's input.
For this setup, we feed the IMU signal into the encoder via the CNN layer, and the historical AU signal, padded with zeros, is input into the decoder for prediction.
Consequently, the model's performance improved from an average AU MAE of 0.267 with the A-Transformer to 0.218, demonstrating the fundamental model skeleton's effectiveness.
However, C-Enc-H's performance is marginally inferior to our H-C-Dec model, with a difference of 0.005 in AU MAE.
This is primarily attributed to the multi-head self-attention (MHSA) layer implemented in the decoder.
The structure of this MHSA layer is as follows:
\begin{equation}
\mathrm{head}_h = \mathrm{Attention}(QW_h^Q, KW_h^K, VW_h^V) = \mathrm{softmax}\left( \frac{(QW_h^Q)(KW_h^K)^T}{\sqrt{d_k}} \right)VW_h^V
\end{equation}
where $W_h^Q$, $W_h^K$, $W_h^V$ are trainable projection matrices for queries, keys and values, and $\sqrt{d_k}$ is the scaling factor.
In the H-C-Dec model, the decoder's cross-attention layer uses queries derived from IMU information, with keys and values sourced from historical AU information in the encoder.
In contrast, the C-Enc-H model primarily focuses on the direct correlations between historical information and current IMU input to process the similarities between historical AU and IMU signals.
The H-C-Dec model, which explores correlations within the feature spaces of IMU and historical AU, fails to capture their true relationship accurately.
\begin{table}[h]
\centering
\caption{Comparison result of different methods}
\label{table:Ablation}
\begin{tabular}{c c c}
\hline
\textbf{Method} & \textbf{Average MAE} \\ \hline
C-Enc & 0.363 \\
A-Transformer & 0.267 \\
C-Enc-H & 0.218 \\ \hline
H-C-Dec(ours) & 0.213\\ \hline
\end{tabular}

\end{table}

\subsection{Micro-benchmark Evaluation}

% 说明采用au mae评估
% \subsubsection{Single vs. Dual IMU}
% To enhance user comfort, as mentioned in Session~\ref{hardware_design}, AUGlasses are equipped with an IMU sensor at both the left and right ends.
% Considering the symmetrical expressions of the 14 AUs, we investigated whether a single sensor can still maintain good performance in face perception.
% For this purpose, we used data from only one IMU to retrain the existing model.
% Due to~\ref{fig:singleimu} the high cost of cross-user training, we conducted validation in two other settings.
% Figure shows that the performance of using two IMUs (L+R) versus a single IMU (L or R) is comparable, with an average difference of just 0.014 (L) and 0.003 (R).
% Specifically, for within-user performance, the average MAE for the left and right IMUs was 0.172 (STD=0.026) and 0.165 (STD=0.030) respectively.
% % For cross-session performance, the average MAE of the left and right IMUs was 0.214 (STD=0.034) and 0.208 (STD=0.032), respectively.
% These findings suggest that a single IMU retains the capability to consistently perceive facial movements.

\subsubsection{Impact of Head Movement}
We conducted a post experiment to understand whether our system can still accurately reconstruct the faces when the head is moving. 
Using the same experiment procedure as that in the previous user study, we asked 3 participants to carry out the entire experiment procedure while walking in a spacious conference room.
They were allowed to move freely around the room to mimic real-life scenarios, where individuals walk at normal speeds and often turn their heads.
During the experiment, participants wore a chest harness that held a smartphone to gather video data from the front.
The entire experiment collected a total of 720 seconds of data (11AU x 10 times * 4 seconds + 7 expressions x 10 times x 4 seconds), and then used the previously set model for inference.

Table~\ref{table:walking} compares the predicted MAE of 14 AUs for the same three participants in walking and previous sitting scenarios.
The performance degradation is mainly due to the deviations in the mapping space of affine transformation.
AUs with weak strength will be masked by artifact signals sometimes.
Even with the degraded performance, our system is still able to reconstruct user faces relatively accurately. 
%Since participants walked throughout the experiment, we relied solely on the initial stationary state for artifact removal, potentially allowing subsequent IMU signals to be re-contaminated with noise.
In real-world scenarios, AUGlasses can detect the user’s stationary state in real time and update mapping parameters to maintain accurate alignment within the mapping space.
This will improve the system performance. 
\begin{table}[h]
\centering
\caption{Performance of AUGlasses in walking and sitting}
\label{table:walking}
\begin{tabular}{|c|c|c|c|c|}
\hline
\multirow{2}{*}{\centering \textbf{Sessions}} & \multirow{2}{*}{\centering \textbf{Scenarios}} & \multicolumn{3}{c|}{\textbf{MAE}}  \\ \cline{3-5} 
                                      &                                     & \textbf{P1}     & \textbf{P4}     & \textbf{P13}     \\ \hline
\multirow{2}{*}{Within User}       & Sitting                             & 0.236  & 0.191  & 0.241  \\ \cline{2-5} 
                                      & Walking                             & 0.416  & 0.429  & 0.413  \\ \hline
\multirow{2}{*}{Cross User}          & Sitting                             & 0.162  & 0.168  & 0.220  \\ \cline{2-5} 
                                      & Walking                             & 0.257  & 0.373  & 0.325  \\ \hline
\end{tabular}

\end{table}
\subsubsection{Long-term Continuous Forecasting including Fine-tuned Cross-user Model}
Our system uses both AU intensities from the previous frame and the IMU data from the current frame for prediction.
Thus the errors from both the AU prediction from the previous frame and the IMU drift will accumulate overtime, which may leads to catastrophic failures. 
Figure~\ref{fig:micro_benchmark}b shows the error variation of the model with different prediction time lengths.
It can be observed that regardless of the setting, the prediction error increases over time.
However, the prediction error stabilizes after 180 seconds.
One explanation for such a behavior is that our system resets the AU intensities to all zeros when they are below a threshold to get rid of accumulated prediction error. 
The IMU data is also remapped each time the head is still to mitigate impact from sensor drift. 
Specifically, the average MAE in the cross-user setting is 0.46 (STD = 0.064), while the average MAE in the within-user setting is 0.36 (STD = 0.058).
Although the model's performance is generally acceptable across different users, there is still potential for optimization on specific individual data.
Therefore, we use data of two repetitions from one session (2016 data samples, about 1.1 minutes) of the test user to fine-tune the cross-user model, enhancing its adaptability to individual user characteristics.
The results show that the average MAE for the fine-tuned model decreased from 0.46 to 0.35, with a standard deviation of 0.053, slightly surpassing the within-user (cross-session) model.
These findings affirm the system's consistently predictable performance over long time duration.

\section{Discussion}

\subsection{Generalizability}
The within-user (cross-session) and cross-user results of AU estimations clearly shows that our model can effectively generalize across different users.
This indicates that our system is not sensitive to slight offsets of sensor contact point, various face widths, and different size and distribution of facial muscles.
There are more ways to improve the generalizability of our system.
1. Our 3D printed glasses frame has plenty room for improvement. 
The glasses can adapt to different face sizes and be more comfortable if the frame is thinner with more flexible materials like metal. 
Properly tuned hinges will also enable a comfortable yet tight contact of the sensor to the skin. 
The design of the sensor supporting structure, especially the housing compartment on the glasses legs, can be further optimized to enable a wider movement range of the contact point. 
2. A more advanced calibration process will greatly help the generalizability of our model. 
The current calibration process only ensures that the signal-to-noise ratio is above a certain threshold. 
It is not able to determine if the contact point is correct or not, which may lead to prediction errors. 
In the future, we plan to develop a more comprehensive calibration process. 
By asking the user to perform various expressions during calibration, it is possible to determine whether the IMU is placed at the right place, or at least at the same spot on the face when the glasses are taken down and worn again. 
In this way, the cross-user prediction accuracy will be improved, especially when personalized data is collected. 
%Even though our 3D printed prototype can fit different face shapes and sizes within a certain range, it is still not as flexible and comfortable as commercial glasses frames using metal material. 

%Although our glasses are designed with adjustable nose pads to fit various nose shapes, cost constraints prevent us from offering multiple sizes for different face shapes.
%Therefore, during the experiment, we excluded participants with abnormally large or small head sizes to protect the glasses and ensure effective sensor performance.
%Nonetheless, variations in head shape still results in the protrusion on AUGlasses being compressed to varying degrees in their initial state, which may affect the accuracy and consistency of the experimental data.
%In section~\ref{cross_user}, we demonstrate that despite these variations, AUGlasses consistently deliver expected outcomes across different users.
%This outcome confirms the device's adaptability and its ability despite variations in user head shapes.
%In the future, we aim to develop commercial glasses that offer greater customization. These eyeglasses will incorporate advanced sensor technology and feature a uniform protrusion design to minimize variations among users.
%We believe that these improvements will significantly enhance both cross-user functionality and the user experience will be greatly improved (Section~\ref{subjectiveresult}).
\subsection{AU Correlation}
Although the Facial Action Coding System (FACS) decomposes facial behavior into 46 distinct motor units, it is challenging for participants to express individual Action Units (AUs) in isolation, whether spontaneously or intentionally. For example, expressing the Nose Wrinkler (AU9) often unintentionally activates the Lid Tightener (AU7). This exemplifies the interconnectedness of facial muscles, where activation of one can inadvertently trigger others. Similarly, some AUs naturally inhibit each other due to opposing muscle movements, such as the Upper Lid Raiser (AU5) and Lid Tightener (AU7), which cannot function simultaneously.

Incorporating knowledge of these AU inter-dependencies into predictive models significantly enhances system performance. By understanding and utilizing these correlations, as demonstrated in various studies~\cite{miriam_jacob_facial_2021, tong_facial_2007, cui_knowledge_2020}, a system can accurately reconstruct facial expressions from a limited number of predicted AUs using minimal hardware like two inertial measurement units (IMUs). Personalizing these models to account for individual variations in facial muscle activation can further improve accuracy, making the distillation of AU correlation knowledge into the model a crucial strategy for advancing facial analysis technologies.

\subsection{User-specific Fine-tune Strategy}
As shown in the previous result, cross-user model performance can significantly benefit from a user-specific fine-tuning phase. 
Currently, the user-specific data used for fine-tuning is equivalent to only 67 seconds of data.
Collecting more data will improve the fine-tuned model performance, but will also increase the user efforts. 
One obvious way to speed up the data collection is to ask users perform a wide variety of compound expressions that involve as many AUs as possible. 
Another ways to improve the data collection efficiency is to focus on those expressions involving AUs that are typically predicted with lower accuracy.
In this way, the model can more rapidly adjust to the individual's unique facial dynamics. 
For our model, AUs have relatively poorer performances are AU7 (Lid Tightener), AU25 (Lips Part), and AU26 (Jaw Drop).
So the expression flow for fine-tune data collection can start from a neutral expression, and gradually move through a flow of expressions that include surprise, fear, and anger. 
Specifically focusing on transitioning smoothly to highlight AU25 (Lips Part) and AU26 (Jaw Drop) during surprise and fear, and AU7 (Lid Tightener) during anger.
Consequently, by asking users to engage in expressive behaviors that challenge the model's weak points, the fine-tuning phase becomes more efficient, leading to improved performance with less data and time investment.

\section{Limitations and Future Work}
\subsection{Potential Contact Spot Drift}
Our prototype uses an adhesive silicone EcoFlex-Gel2 to ensure stable contact to the skin.
However, the silicone is hydrophobic, which means it will lose adhesiveness when there is liquid on the skin contact point. 
This usually happens when the user sweats or just washed their face. 
Materials like hydrogel can alleviate the issue and remain adhesive under such circumstances. 
We plan to test different materials in the future to find material that can work under different scenarios in the future.

%Even though our 3D printed prototype can fit different face shapes and sizes within a certain range, it is still not as flexible and comfortable as commercial glasses frames using metal material. 

%Although our glasses are designed with adjustable nose pads to fit various nose shapes, cost constraints prevent us from offering multiple sizes for different face shapes.
%Therefore, during the experiment, we excluded participants with abnormally large or small head sizes to protect the glasses and ensure effective sensor performance.
%Nonetheless, variations in head shape still results in the protrusion on AUGlasses being compressed to varying degrees in their initial state, which may affect the accuracy and consistency of the experimental data.
%In section~\ref{cross_user}, we demonstrate that despite these variations, AUGlasses consistently deliver expected outcomes across different users.
%This outcome confirms the device's adaptability and its ability despite variations in user head shapes.
%In the future, we aim to develop commercial glasses that offer greater customization. These eyeglasses will incorporate advanced sensor technology and feature a uniform protrusion design to minimize variations among users.
%We believe that these improvements will significantly enhance both cross-user functionality and the user experience will be greatly improved (Section~\ref{subjectiveresult}).

\subsection{Blink Detection and Handling}
Currently, our system employs a filter during the facial reconstruction process since facial movements are continuous. 
It is inevitable that blinks will interfere with our system since it causes large skin movements around the temporal area. 
One way to address the issue is the add a binary blink detector based on IMU signals.
The frames when blink is detected can be discarded and only those without blinks will start the inference process. 
The AU intensities of those blink frames can be replaced with interpolated results from non-blink frames before and after. 
This will still generate truthful facial reconstruction since our system has a high frame rate of 30 fps, which can tolerate frame losses. 
We plan to implement such a blink detector in our system to enable more robust facial reconstruction with accurate blink detection. 

\section{Conclusion}
We introduce AUGlasses, a low-power and unobtrusive smart glasses that continuously track 14 facial action units for accurate facial reconstruction. 
We design a novel real-time AU estimation transformer architecture, which utilizes IMU data and previous AU intensities to estimate AU strengths of the current frame. 
We then reconstruct an avatar's face in Unity based on the predicted strengths of the 14 AUs, and compared the facial reconstruction performance with ground truth by extracting 51 facial landmarks on the face. 
The MAE for the reconstructed 3D face was only 1.93 mm, with a NME of 2.75\%.
Subjective results show more than 70\% of participants thought AUGlasses were comfortable to wear and suitable for daily use.
Post analysis show that our system can work with a single IMU with a 50Hz sampling rate, which makes it possible further reduces the power consumption to 40.96 mW. 
% In the more challenging situation of walking scenarios, AUGlasses still perform well with a cross-user MAE of xx. \todo{}
We believe AUGlasses is more practical solution for facial reconstruction on smart glasses, opening up possibilities for various exciting applications like virtual social interaction and emotion awareness.
\begin{acks}
To Robert, for the bagels and explaining CMYK and color spaces.
\end{acks}

\bibliographystyle{ACM-Reference-Format}
\bibliography{references}


\begin{thebibliography}{69}

%%% the Web address.

\ifx \showCODEN    \undefined \def \showCODEN     #1{\unskip}     \fi
\ifx \showDOI      \undefined \def \showDOI       #1{#1}\fi
\ifx \showISBNx    \undefined \def \showISBNx     #1{\unskip}     \fi
\ifx \showISBNxiii \undefined \def \showISBNxiii  #1{\unskip}     \fi
\ifx \showISSN     \undefined \def \showISSN      #1{\unskip}     \fi
\ifx \showLCCN     \undefined \def \showLCCN      #1{\unskip}     \fi
\ifx \shownote     \undefined \def \shownote      #1{#1}          \fi
\ifx \showarticletitle \undefined \def \showarticletitle #1{#1}   \fi
\ifx \showURL      \undefined \def \showURL       {\relax}        \fi
% invisible to TeX
\providecommand\bibfield[2]{#2}
\providecommand\bibinfo[2]{#2}
\providecommand\natexlab[1]{#1}
\providecommand\showeprint[2][]{arXiv:#2}

\bibitem[Aoki et~al\mbox{.}(2021)]%
        {aoki_facerecglasses_2021}
\bibfield{author}{\bibinfo{person}{Hiroaki Aoki}, \bibinfo{person}{Ayumi Ohnishi}, \bibinfo{person}{Naoya Isoyama}, \bibinfo{person}{Tsutomu Terada}, {and} \bibinfo{person}{Masahiko Tsukamoto}.} \bibinfo{year}{2021}\natexlab{}.
\newblock \showarticletitle{{FaceRecGlasses}: {A} {Wearable} {System} for {Recognizing} {Self} {Facial} {Expressions} {Using} {Compact} {Wearable} {Cameras}}. In \bibinfo{booktitle}{\emph{Proceedings of the {Augmented} {Humans} {International} {Conference} 2021}} \emph{(\bibinfo{series}{{AHs} '21})}. \bibinfo{publisher}{Association for Computing Machinery}, \bibinfo{address}{New York, NY, USA}, \bibinfo{pages}{55--65}.
\newblock
\showISBNx{978-1-4503-8428-5}
\urldef\tempurl%
\url{https://doi.org/10.1145/3458709.3458983}
\showDOI{\tempurl}


\bibitem[Argaud et~al\mbox{.}(2018)]%
        {argaud_facial_2018}
\bibfield{author}{\bibinfo{person}{Soizic Argaud}, \bibinfo{person}{Marc Vérin}, \bibinfo{person}{Paul Sauleau}, {and} \bibinfo{person}{Didier Grandjean}.} \bibinfo{year}{2018}\natexlab{}.
\newblock \showarticletitle{Facial emotion recognition in {Parkinson}'s disease: {A} review and new hypotheses}.
\newblock \bibinfo{journal}{\emph{Movement Disorders: Official Journal of the Movement Disorder Society}} \bibinfo{volume}{33}, \bibinfo{number}{4} (\bibinfo{date}{April} \bibinfo{year}{2018}), \bibinfo{pages}{554--567}.
\newblock
\showISSN{1531-8257}
\urldef\tempurl%
\url{https://doi.org/10.1002/mds.27305}
\showDOI{\tempurl}


\bibitem[Baltrušaitis et~al\mbox{.}(2016)]%
        {baltrusaitis_openface_2016}
\bibfield{author}{\bibinfo{person}{Tadas Baltrušaitis}, \bibinfo{person}{Peter Robinson}, {and} \bibinfo{person}{Louis-Philippe Morency}.} \bibinfo{year}{2016}\natexlab{}.
\newblock \showarticletitle{{OpenFace}: {An} open source facial behavior analysis toolkit}. In \bibinfo{booktitle}{\emph{2016 {IEEE} {Winter} {Conference} on {Applications} of {Computer} {Vision} ({WACV})}}. \bibinfo{pages}{1--10}.
\newblock
\urldef\tempurl%
\url{https://doi.org/10.1109/WACV.2016.7477553}
\showDOI{\tempurl}


\bibitem[Bassili(1979)]%
        {bassili_emotion_1979}
\bibfield{author}{\bibinfo{person}{J.~N. Bassili}.} \bibinfo{year}{1979}\natexlab{}.
\newblock \showarticletitle{Emotion recognition: the role of facial movement and the relative importance of upper and lower areas of the face}.
\newblock \bibinfo{journal}{\emph{Journal of Personality and Social Psychology}} \bibinfo{volume}{37}, \bibinfo{number}{11} (\bibinfo{date}{Nov.} \bibinfo{year}{1979}), \bibinfo{pages}{2049--2058}.
\newblock
\showISSN{0022-3514}
\urldef\tempurl%
\url{https://doi.org/10.1037//0022-3514.37.11.2049}
\showDOI{\tempurl}


\bibitem[Bedri et~al\mbox{.}(2020)]%
        {bedri_fitbyte_2020}
\bibfield{author}{\bibinfo{person}{Abdelkareem Bedri}, \bibinfo{person}{Diana Li}, \bibinfo{person}{Rushil Khurana}, \bibinfo{person}{Kunal Bhuwalka}, {and} \bibinfo{person}{Mayank Goel}.} \bibinfo{year}{2020}\natexlab{}.
\newblock \showarticletitle{{FitByte}: {Automatic} {Diet} {Monitoring} in {Unconstrained} {Situations} {Using} {Multimodal} {Sensing} on {Eyeglasses}}. In \bibinfo{booktitle}{\emph{Proceedings of the 2020 {CHI} {Conference} on {Human} {Factors} in {Computing} {Systems}}} \emph{(\bibinfo{series}{{CHI} '20})}. \bibinfo{publisher}{Association for Computing Machinery}, \bibinfo{address}{New York, NY, USA}, \bibinfo{pages}{1--12}.
\newblock
\showISBNx{978-1-4503-6708-0}
\urldef\tempurl%
\url{https://doi.org/10.1145/3313831.3376869}
\showDOI{\tempurl}


\bibitem[Bedri et~al\mbox{.}(2017)]%
        {bedri_earbit_2017}
\bibfield{author}{\bibinfo{person}{Abdelkareem Bedri}, \bibinfo{person}{Richard Li}, \bibinfo{person}{Malcolm Haynes}, \bibinfo{person}{Raj~Prateek Kosaraju}, \bibinfo{person}{Ishaan Grover}, \bibinfo{person}{Temiloluwa Prioleau}, \bibinfo{person}{Min~Yan Beh}, \bibinfo{person}{Mayank Goel}, \bibinfo{person}{Thad Starner}, {and} \bibinfo{person}{Gregory Abowd}.} \bibinfo{year}{2017}\natexlab{}.
\newblock \showarticletitle{{EarBit}: {Using} {Wearable} {Sensors} to {Detect} {Eating} {Episodes} in {Unconstrained} {Environments}}.
\newblock \bibinfo{journal}{\emph{Proceedings of the ACM on Interactive, Mobile, Wearable and Ubiquitous Technologies}} \bibinfo{volume}{1}, \bibinfo{number}{3} (\bibinfo{year}{2017}), \bibinfo{pages}{37:1--37:20}.
\newblock
\urldef\tempurl%
\url{https://doi.org/10.1145/3130902}
\showDOI{\tempurl}


\bibitem[Bello et~al\mbox{.}(2023)]%
        {bello_inmyface_2023}
\bibfield{author}{\bibinfo{person}{Hymalai Bello}, \bibinfo{person}{Luis Alfredo~Sanchez Marin}, \bibinfo{person}{Sungho Suh}, \bibinfo{person}{Bo Zhou}, {and} \bibinfo{person}{Paul Lukowicz}.} \bibinfo{year}{2023}\natexlab{}.
\newblock \bibinfo{title}{{InMyFace}: {Inertial} and {Mechanomyography}-{Based} {Sensor} {Fusion} for {Wearable} {Facial} {Activity} {Recognition}}.
\newblock
\newblock
\urldef\tempurl%
\url{https://doi.org/10.48550/arXiv.2302.04024}
\showDOI{\tempurl}


\bibitem[Bi et~al\mbox{.}(2018)]%
        {bi_auracle_2018}
\bibfield{author}{\bibinfo{person}{Shengjie Bi}, \bibinfo{person}{Tao Wang}, \bibinfo{person}{Nicole Tobias}, \bibinfo{person}{Josephine Nordrum}, \bibinfo{person}{Shang Wang}, \bibinfo{person}{George Halvorsen}, \bibinfo{person}{Sougata Sen}, \bibinfo{person}{Ronald Peterson}, \bibinfo{person}{Kofi Odame}, \bibinfo{person}{Kelly Caine}, \bibinfo{person}{Ryan Halter}, \bibinfo{person}{Jacob Sorber}, {and} \bibinfo{person}{David Kotz}.} \bibinfo{year}{2018}\natexlab{}.
\newblock \showarticletitle{Auracle: {Detecting} {Eating} {Episodes} with an {Ear}-mounted {Sensor}}.
\newblock \bibinfo{journal}{\emph{Proceedings of the ACM on Interactive, Mobile, Wearable and Ubiquitous Technologies}} \bibinfo{volume}{2}, \bibinfo{number}{3} (\bibinfo{year}{2018}), \bibinfo{pages}{92:1--92:27}.
\newblock
\urldef\tempurl%
\url{https://doi.org/10.1145/3264902}
\showDOI{\tempurl}


\bibitem[Cao et~al\mbox{.}(2015)]%
        {cao_real-time_2015}
\bibfield{author}{\bibinfo{person}{Chen Cao}, \bibinfo{person}{Derek Bradley}, \bibinfo{person}{Kun Zhou}, {and} \bibinfo{person}{Thabo Beeler}.} \bibinfo{year}{2015}\natexlab{}.
\newblock \showarticletitle{Real-time high-fidelity facial performance capture}.
\newblock \bibinfo{journal}{\emph{ACM Transactions on Graphics}} \bibinfo{volume}{34}, \bibinfo{number}{4} (\bibinfo{year}{2015}), \bibinfo{pages}{46:1--46:9}.
\newblock
\showISSN{0730-0301}
\urldef\tempurl%
\url{https://doi.org/10.1145/2766943}
\showDOI{\tempurl}


\bibitem[Cha et~al\mbox{.}(2020)]%
        {cha_real-time_2020}
\bibfield{author}{\bibinfo{person}{Ho-Seung Cha}, \bibinfo{person}{Seong-Jun Choi}, {and} \bibinfo{person}{Chang-Hwan Im}.} \bibinfo{year}{2020}\natexlab{}.
\newblock \showarticletitle{Real-{Time} {Recognition} of {Facial} {Expressions} {Using} {Facial} {Electromyograms} {Recorded} {Around} the {Eyes} for {Social} {Virtual} {Reality} {Applications}}.
\newblock \bibinfo{journal}{\emph{IEEE Access}}  \bibinfo{volume}{8} (\bibinfo{year}{2020}), \bibinfo{pages}{62065--62075}.
\newblock
\showISSN{2169-3536}
\urldef\tempurl%
\url{https://doi.org/10.1109/ACCESS.2020.2983608}
\showDOI{\tempurl}


\bibitem[Chen et~al\mbox{.}(2021)]%
        {chen_neckface_2021}
\bibfield{author}{\bibinfo{person}{Tuochao Chen}, \bibinfo{person}{Yaxuan Li}, \bibinfo{person}{Songyun Tao}, \bibinfo{person}{Hyunchul Lim}, \bibinfo{person}{Mose Sakashita}, \bibinfo{person}{Ruidong Zhang}, \bibinfo{person}{Francois Guimbretiere}, {and} \bibinfo{person}{Cheng Zhang}.} \bibinfo{year}{2021}\natexlab{}.
\newblock \showarticletitle{{NeckFace}: {Continuously} {Tracking} {Full} {Facial} {Expressions} on {Neck}-mounted {Wearables}}.
\newblock \bibinfo{journal}{\emph{Proceedings of the ACM on Interactive, Mobile, Wearable and Ubiquitous Technologies}} \bibinfo{volume}{5}, \bibinfo{number}{2} (\bibinfo{year}{2021}), \bibinfo{pages}{58:1--58:31}.
\newblock
\urldef\tempurl%
\url{https://doi.org/10.1145/3463511}
\showDOI{\tempurl}


\bibitem[Chen et~al\mbox{.}(2020)]%
        {chen_c-face_2020}
\bibfield{author}{\bibinfo{person}{Tuochao Chen}, \bibinfo{person}{Benjamin Steeper}, \bibinfo{person}{Kinan Alsheikh}, \bibinfo{person}{Songyun Tao}, \bibinfo{person}{François Guimbretière}, {and} \bibinfo{person}{Cheng Zhang}.} \bibinfo{year}{2020}\natexlab{}.
\newblock \showarticletitle{C-{Face}: {Continuously} {Reconstructing} {Facial} {Expressions} by {Deep} {Learning} {Contours} of the {Face} with {Ear}-mounted {Miniature} {Cameras}}. In \bibinfo{booktitle}{\emph{Proceedings of the 33rd {Annual} {ACM} {Symposium} on {User} {Interface} {Software} and {Technology}}} \emph{(\bibinfo{series}{{UIST} '20})}. \bibinfo{publisher}{Association for Computing Machinery}, \bibinfo{address}{New York, NY, USA}, \bibinfo{pages}{112--125}.
\newblock
\showISBNx{978-1-4503-7514-6}
\urldef\tempurl%
\url{https://doi.org/10.1145/3379337.3415879}
\showDOI{\tempurl}


\bibitem[Choi et~al\mbox{.}(2022)]%
        {choi_ppgface_2022}
\bibfield{author}{\bibinfo{person}{Seokmin Choi}, \bibinfo{person}{Yang Gao}, \bibinfo{person}{Yincheng Jin}, \bibinfo{person}{Se~jun Kim}, \bibinfo{person}{Jiyang Li}, \bibinfo{person}{Wenyao Xu}, {and} \bibinfo{person}{Zhanpeng Jin}.} \bibinfo{year}{2022}\natexlab{}.
\newblock \showarticletitle{{PPGface}: {Like} {What} {You} {Are} {Watching}? {Earphones} {Can} "{Feel}" {Your} {Facial} {Expressions}}.
\newblock \bibinfo{journal}{\emph{Proceedings of the ACM on Interactive, Mobile, Wearable and Ubiquitous Technologies}} \bibinfo{volume}{6}, \bibinfo{number}{2} (\bibinfo{year}{2022}), \bibinfo{pages}{48:1--48:32}.
\newblock
\urldef\tempurl%
\url{https://doi.org/10.1145/3534597}
\showDOI{\tempurl}


\bibitem[Cui et~al\mbox{.}(2020)]%
        {cui_knowledge_2020}
\bibfield{author}{\bibinfo{person}{Zijun Cui}, \bibinfo{person}{Tengfei Song}, \bibinfo{person}{Yuru Wang}, {and} \bibinfo{person}{Qiang Ji}.} \bibinfo{year}{2020}\natexlab{}.
\newblock \showarticletitle{Knowledge {Augmented} {Deep} {Neural} {Networks} for {Joint} {Facial} {Expression} and {Action} {Unit} {Recognition}}. In \bibinfo{booktitle}{\emph{Advances in {Neural} {Information} {Processing} {Systems}}}, Vol.~\bibinfo{volume}{33}. \bibinfo{publisher}{Curran Associates, Inc.}, \bibinfo{pages}{14338--14349}.
\newblock
\urldef\tempurl%
\url{https://proceedings.neurips.cc/paper_files/paper/2020/hash/a51fb975227d6640e4fe47854476d133-Abstract.html}
\showURL{%
\tempurl}


\bibitem[Danieau et~al\mbox{.}(2019)]%
        {danieau_automatic_2019}
\bibfield{author}{\bibinfo{person}{Fabien Danieau}, \bibinfo{person}{Ilja Gubins}, \bibinfo{person}{Nicolas Olivier}, \bibinfo{person}{Olivier Dumas}, \bibinfo{person}{Bernard Denis}, \bibinfo{person}{Thomas Lopez}, \bibinfo{person}{Nicolas Mollet}, \bibinfo{person}{Brian Frager}, {and} \bibinfo{person}{Quentin Avril}.} \bibinfo{year}{2019}\natexlab{}.
\newblock \showarticletitle{Automatic {Generation} and {Stylization} of {3D} {Facial} {Rigs}}. In \bibinfo{booktitle}{\emph{2019 {IEEE} {Conference} on {Virtual} {Reality} and {3D} {User} {Interfaces} ({VR})}}. \bibinfo{pages}{784--792}.
\newblock
\urldef\tempurl%
\url{https://doi.org/10.1109/VR.2019.8798208}
\showDOI{\tempurl}
\newblock
\shownote{ISSN: 2642-5254}.


\bibitem[Dong et~al\mbox{.}(2018)]%
        {dong_style_2018}
\bibfield{author}{\bibinfo{person}{Xuanyi Dong}, \bibinfo{person}{Yan Yan}, \bibinfo{person}{Wanli Ouyang}, {and} \bibinfo{person}{Yi Yang}.} \bibinfo{year}{2018}\natexlab{}.
\newblock \showarticletitle{Style {Aggregated} {Network} for {Facial} {Landmark} {Detection}}. In \bibinfo{booktitle}{\emph{2018 {IEEE}/{CVF} {Conference} on {Computer} {Vision} and {Pattern} {Recognition}}}. \bibinfo{publisher}{IEEE}, \bibinfo{address}{Salt Lake City, UT, USA}, \bibinfo{pages}{379--388}.
\newblock
\showISBNx{978-1-5386-6420-9}
\urldef\tempurl%
\url{https://doi.org/10.1109/CVPR.2018.00047}
\showDOI{\tempurl}


\bibitem[Ekman and Friesen(1978)]%
        {ekman_facial_1978}
\bibfield{author}{\bibinfo{person}{Paul Ekman} {and} \bibinfo{person}{Wallace~V Friesen}.} \bibinfo{year}{1978}\natexlab{}.
\newblock \showarticletitle{Facial action coding system}.
\newblock \bibinfo{journal}{\emph{Environmental Psychology \& Nonverbal Behavior}} (\bibinfo{year}{1978}).
\newblock


\bibitem[Frith(2009)]%
        {frith_role_2009}
\bibfield{author}{\bibinfo{person}{Chris Frith}.} \bibinfo{year}{2009}\natexlab{}.
\newblock \showarticletitle{Role of facial expressions in social interactions}.
\newblock \bibinfo{journal}{\emph{Philosophical Transactions of the Royal Society of London. Series B, Biological Sciences}} \bibinfo{volume}{364}, \bibinfo{number}{1535} (\bibinfo{date}{Dec.} \bibinfo{year}{2009}), \bibinfo{pages}{3453--3458}.
\newblock
\showISSN{1471-2970}
\urldef\tempurl%
\url{https://doi.org/10.1098/rstb.2009.0142}
\showDOI{\tempurl}


\bibitem[Guo and Liang(2023)]%
        {guo_texonmask_2023}
\bibfield{author}{\bibinfo{person}{Zengrong Guo} {and} \bibinfo{person}{Rong-Hao Liang}.} \bibinfo{year}{2023}\natexlab{}.
\newblock \showarticletitle{{TexonMask}: {Facial} {Expression} {Recognition} {Using} {Textile} {Electrodes} on {Commodity} {Facemasks}}. In \bibinfo{booktitle}{\emph{Proceedings of the 2023 {CHI} {Conference} on {Human} {Factors} in {Computing} {Systems}}} \emph{(\bibinfo{series}{{CHI} '23})}. \bibinfo{publisher}{Association for Computing Machinery}, \bibinfo{address}{New York, NY, USA}, \bibinfo{pages}{1--15}.
\newblock
\showISBNx{978-1-4503-9421-5}
\urldef\tempurl%
\url{https://doi.org/10.1145/3544548.3581295}
\showDOI{\tempurl}


\bibitem[Kar et~al\mbox{.}(2024)]%
        {kar_fiducial_2024}
\bibfield{author}{\bibinfo{person}{Purbayan Kar}, \bibinfo{person}{Vishal Chudasama}, \bibinfo{person}{Naoyuki Onoe}, \bibinfo{person}{Pankaj Wasnik}, {and} \bibinfo{person}{Vineeth Balasubramanian}.} \bibinfo{year}{2024}\natexlab{}.
\newblock \bibinfo{title}{Fiducial {Focus} {Augmentation} for {Facial} {Landmark} {Detection}}.
\newblock
\newblock
\urldef\tempurl%
\url{https://doi.org/10.48550/arXiv.2402.15044}
\showDOI{\tempurl}
\newblock
\shownote{arXiv:2402.15044 [cs]}.


\bibitem[Kim and Brienza(2006)]%
        {kim_development_2006}
\bibfield{author}{\bibinfo{person}{Jong~Bae Kim} {and} \bibinfo{person}{David~M. Brienza}.} \bibinfo{year}{2006}\natexlab{}.
\newblock \showarticletitle{Development of a {Remote} {Accessibility} {Assessment} {System} through three-dimensional reconstruction technology}.
\newblock \bibinfo{journal}{\emph{Journal of Rehabilitation Research and Development}} \bibinfo{volume}{43}, \bibinfo{number}{2} (\bibinfo{year}{2006}), \bibinfo{pages}{257--272}.
\newblock
\showISSN{1938-1352}
\urldef\tempurl%
\url{https://doi.org/10.1682/jrrd.2004.12.0163}
\showDOI{\tempurl}


\bibitem[Krumhuber et~al\mbox{.}(2023)]%
        {krumhuber_role_2023}
\bibfield{author}{\bibinfo{person}{Eva~G. Krumhuber}, \bibinfo{person}{Lina~I. Skora}, \bibinfo{person}{Harold C.~H. Hill}, {and} \bibinfo{person}{Karen Lander}.} \bibinfo{year}{2023}\natexlab{}.
\newblock \showarticletitle{The role of facial movements in emotion recognition}.
\newblock \bibinfo{journal}{\emph{Nature Reviews Psychology}} \bibinfo{volume}{2}, \bibinfo{number}{5} (\bibinfo{date}{May} \bibinfo{year}{2023}), \bibinfo{pages}{283--296}.
\newblock
\showISSN{2731-0574}
\urldef\tempurl%
\url{https://doi.org/10.1038/s44159-023-00172-1}
\showDOI{\tempurl}
\newblock
\shownote{Publisher: Nature Publishing Group}.


\bibitem[Kwapisz et~al\mbox{.}(2011)]%
        {kwapisz_activity_2011}
\bibfield{author}{\bibinfo{person}{Jennifer~R. Kwapisz}, \bibinfo{person}{Gary~M. Weiss}, {and} \bibinfo{person}{Samuel~A. Moore}.} \bibinfo{year}{2011}\natexlab{}.
\newblock \showarticletitle{Activity recognition using cell phone accelerometers}.
\newblock \bibinfo{journal}{\emph{ACM SIGKDD Explorations Newsletter}} \bibinfo{volume}{12}, \bibinfo{number}{2} (\bibinfo{year}{2011}), \bibinfo{pages}{74--82}.
\newblock
\showISSN{1931-0145}
\urldef\tempurl%
\url{https://doi.org/10.1145/1964897.1964918}
\showDOI{\tempurl}


\bibitem[Lapatki et~al\mbox{.}(2003)]%
        {lapatki_surface_2003}
\bibfield{author}{\bibinfo{person}{B.~G. Lapatki}, \bibinfo{person}{D.~F. Stegeman}, {and} \bibinfo{person}{I.~E. Jonas}.} \bibinfo{year}{2003}\natexlab{}.
\newblock \showarticletitle{A surface {EMG} electrode for the simultaneous observation of multiple facial muscles}.
\newblock \bibinfo{journal}{\emph{Journal of Neuroscience Methods}} \bibinfo{volume}{123}, \bibinfo{number}{2} (\bibinfo{date}{March} \bibinfo{year}{2003}), \bibinfo{pages}{117--128}.
\newblock
\showISSN{0165-0270}
\urldef\tempurl%
\url{https://doi.org/10.1016/s0165-0270(02)00323-0}
\showDOI{\tempurl}


\bibitem[Leblond et~al\mbox{.}(2021)]%
        {leblond_machine_2021}
\bibfield{author}{\bibinfo{person}{Rémi Leblond}, \bibinfo{person}{Jean-Baptiste Alayrac}, \bibinfo{person}{Laurent Sifre}, \bibinfo{person}{Miruna Pislar}, \bibinfo{person}{Lespiau Jean-Baptiste}, \bibinfo{person}{Ioannis Antonoglou}, \bibinfo{person}{Karen Simonyan}, {and} \bibinfo{person}{Oriol Vinyals}.} \bibinfo{year}{2021}\natexlab{}.
\newblock \showarticletitle{Machine {Translation} {Decoding} beyond {Beam} {Search}}. In \bibinfo{booktitle}{\emph{Proceedings of the 2021 {Conference} on {Empirical} {Methods} in {Natural} {Language} {Processing}}}, \bibfield{editor}{\bibinfo{person}{Marie-Francine Moens}, \bibinfo{person}{Xuanjing Huang}, \bibinfo{person}{Lucia Specia}, {and} \bibinfo{person}{Scott Wen-tau Yih}} (Eds.). \bibinfo{publisher}{Association for Computational Linguistics}, \bibinfo{address}{Online and Punta Cana, Dominican Republic}, \bibinfo{pages}{8410--8434}.
\newblock
\urldef\tempurl%
\url{https://doi.org/10.18653/v1/2021.emnlp-main.662}
\showDOI{\tempurl}


\bibitem[Li et~al\mbox{.}(2022)]%
        {li_eario_2022}
\bibfield{author}{\bibinfo{person}{Ke Li}, \bibinfo{person}{Ruidong Zhang}, \bibinfo{person}{Bo Liang}, \bibinfo{person}{François Guimbretière}, {and} \bibinfo{person}{Cheng Zhang}.} \bibinfo{year}{2022}\natexlab{}.
\newblock \showarticletitle{{EarIO}: {A} {Low}-power {Acoustic} {Sensing} {Earable} for {Continuously} {Tracking} {Detailed} {Facial} {Movements}}.
\newblock \bibinfo{journal}{\emph{Proceedings of the ACM on Interactive, Mobile, Wearable and Ubiquitous Technologies}} \bibinfo{volume}{6}, \bibinfo{number}{2} (\bibinfo{date}{July} \bibinfo{year}{2022}), \bibinfo{pages}{1--24}.
\newblock
\showISSN{2474-9567}
\urldef\tempurl%
\url{https://doi.org/10.1145/3534621}
\showDOI{\tempurl}


\bibitem[Li et~al\mbox{.}(2023)]%
        {li_human_2023}
\bibfield{author}{\bibinfo{person}{Xiaoxiong Li}, \bibinfo{person}{Si Chen}, \bibinfo{person}{Shuning Zhang}, \bibinfo{person}{Linsheng Hou}, \bibinfo{person}{Yuying Zhu}, {and} \bibinfo{person}{Zelong Xiao}.} \bibinfo{year}{2023}\natexlab{}.
\newblock \showarticletitle{Human {Activity} {Recognition} {Using} {IR}-{UWB} {Radar}: {A} {Lightweight} {Transformer} {Approach}}.
\newblock \bibinfo{journal}{\emph{IEEE Geoscience and Remote Sensing Letters}}  \bibinfo{volume}{20} (\bibinfo{year}{2023}), \bibinfo{pages}{1--5}.
\newblock
\showISSN{1558-0571}
\urldef\tempurl%
\url{https://doi.org/10.1109/LGRS.2023.3314628}
\showDOI{\tempurl}
\newblock
\shownote{Conference Name: IEEE Geoscience and Remote Sensing Letters}.


\bibitem[Liang et~al\mbox{.}(2004)]%
        {liang_new_2004}
\bibfield{author}{\bibinfo{person}{Rong-Hua Liang}, \bibinfo{person}{Zhi-Geng Pan}, {and} \bibinfo{person}{Chun Chen}.} \bibinfo{year}{2004}\natexlab{}.
\newblock \showarticletitle{New algorithm for {3D} facial model reconstruction and its application in virtual reality}.
\newblock \bibinfo{journal}{\emph{Journal of Computer Science and Technology}} \bibinfo{volume}{19}, \bibinfo{number}{4} (\bibinfo{date}{July} \bibinfo{year}{2004}), \bibinfo{pages}{501--509}.
\newblock
\showISSN{1860-4749}
\urldef\tempurl%
\url{https://doi.org/10.1007/BF02944751}
\showDOI{\tempurl}


\bibitem[Liu et~al\mbox{.}(2021)]%
        {liu_confidence-aware_2021}
\bibfield{author}{\bibinfo{person}{Yijin Liu}, \bibinfo{person}{Fandong Meng}, \bibinfo{person}{Yufeng Chen}, \bibinfo{person}{Jinan Xu}, {and} \bibinfo{person}{Jie Zhou}.} \bibinfo{year}{2021}\natexlab{}.
\newblock \showarticletitle{Confidence-{Aware} {Scheduled} {Sampling} for {Neural} {Machine} {Translation}}. In \bibinfo{booktitle}{\emph{Findings of the {Association} for {Computational} {Linguistics}: {ACL}-{IJCNLP} 2021}}, \bibfield{editor}{\bibinfo{person}{Chengqing Zong}, \bibinfo{person}{Fei Xia}, \bibinfo{person}{Wenjie Li}, {and} \bibinfo{person}{Roberto Navigli}} (Eds.). \bibinfo{publisher}{Association for Computational Linguistics}, \bibinfo{address}{Online}, \bibinfo{pages}{2327--2337}.
\newblock
\urldef\tempurl%
\url{https://doi.org/10.18653/v1/2021.findings-acl.205}
\showDOI{\tempurl}


\bibitem[Mao et~al\mbox{.}(2010)]%
        {mao_facial_2010}
\bibfield{author}{\bibinfo{person}{Jeremy~J. Mao}, \bibinfo{person}{Michael~S. Stosich}, \bibinfo{person}{Eduardo~K. Moioli}, \bibinfo{person}{Chang~Hun Lee}, \bibinfo{person}{Susan~Y. Fu}, \bibinfo{person}{Barbara Bastian}, \bibinfo{person}{Sidney~B. Eisig}, \bibinfo{person}{Candice Zemnick}, \bibinfo{person}{Jeffrey Ascherman}, \bibinfo{person}{June Wu}, \bibinfo{person}{Christine Rohde}, {and} \bibinfo{person}{Jeffrey Ahn}.} \bibinfo{year}{2010}\natexlab{}.
\newblock \showarticletitle{Facial {Reconstruction} by {Biosurgery}: {Cell} {Transplantation} {Versus} {Cell} {Homing}}.
\newblock \bibinfo{journal}{\emph{Tissue Engineering Part B: Reviews}} \bibinfo{volume}{16}, \bibinfo{number}{2} (\bibinfo{date}{April} \bibinfo{year}{2010}), \bibinfo{pages}{257--262}.
\newblock
\showISSN{1937-3368}
\urldef\tempurl%
\url{https://doi.org/10.1089/ten.teb.2009.0496}
\showDOI{\tempurl}
\newblock
\shownote{Publisher: Mary Ann Liebert, Inc., publishers}.


\bibitem[Martyniuk et~al\mbox{.}(2022)]%
        {martyniuk_dad-3dheads_2022}
\bibfield{author}{\bibinfo{person}{Tetiana Martyniuk}, \bibinfo{person}{Orest Kupyn}, \bibinfo{person}{Yana Kurlyak}, \bibinfo{person}{Igor Krashenyi}, \bibinfo{person}{Jiri Matas}, {and} \bibinfo{person}{Viktoriia Sharmanska}.} \bibinfo{year}{2022}\natexlab{}.
\newblock \showarticletitle{{DAD}-{3DHeads}: {A} {Large}-scale {Dense}, {Accurate} and {Diverse} {Dataset} for {3D} {Head} {Alignment} from a {Single} {Image}}. In \bibinfo{booktitle}{\emph{2022 {IEEE}/{CVF} {Conference} on {Computer} {Vision} and {Pattern} {Recognition} ({CVPR})}}. \bibinfo{publisher}{IEEE}, \bibinfo{address}{New Orleans, LA, USA}, \bibinfo{pages}{20910--20920}.
\newblock
\showISBNx{978-1-66546-946-3}
\urldef\tempurl%
\url{https://doi.org/10.1109/CVPR52688.2022.02027}
\showDOI{\tempurl}


\bibitem[Masai et~al\mbox{.}(2017)]%
        {masai_evaluation_2017}
\bibfield{author}{\bibinfo{person}{Katsutoshi Masai}, \bibinfo{person}{Kai Kunze}, \bibinfo{person}{Yuta Sugiura}, \bibinfo{person}{Masa Ogata}, \bibinfo{person}{Masahiko Inami}, {and} \bibinfo{person}{Maki Sugimoto}.} \bibinfo{year}{2017}\natexlab{}.
\newblock \showarticletitle{Evaluation of {Facial} {Expression} {Recognition} by a {Smart} {Eyewear} for {Facial} {Direction} {Changes}, {Repeatability}, and {Positional} {Drift}}.
\newblock \bibinfo{journal}{\emph{ACM Transactions on Interactive Intelligent Systems}} \bibinfo{volume}{7}, \bibinfo{number}{4} (\bibinfo{date}{Dec.} \bibinfo{year}{2017}), \bibinfo{pages}{1--23}.
\newblock
\showISSN{2160-6455, 2160-6463}
\urldef\tempurl%
\url{https://doi.org/10.1145/3012941}
\showDOI{\tempurl}


\bibitem[Matthies et~al\mbox{.}(2017)]%
        {matthies_earfieldsensing_2017}
\bibfield{author}{\bibinfo{person}{Denys J.~C. Matthies}, \bibinfo{person}{Bernhard~A. Strecker}, {and} \bibinfo{person}{Bodo Urban}.} \bibinfo{year}{2017}\natexlab{}.
\newblock \showarticletitle{{EarFieldSensing}: {A} {Novel} {In}-{Ear} {Electric} {Field} {Sensing} to {Enrich} {Wearable} {Gesture} {Input} through {Facial} {Expressions}}. In \bibinfo{booktitle}{\emph{Proceedings of the 2017 {CHI} {Conference} on {Human} {Factors} in {Computing} {Systems}}} \emph{(\bibinfo{series}{{CHI} '17})}. \bibinfo{publisher}{Association for Computing Machinery}, \bibinfo{address}{New York, NY, USA}, \bibinfo{pages}{1911--1922}.
\newblock
\showISBNx{978-1-4503-4655-9}
\urldef\tempurl%
\url{https://doi.org/10.1145/3025453.3025692}
\showDOI{\tempurl}


\bibitem[Miriam~Jacob and Stenger(2021)]%
        {miriam_jacob_facial_2021}
\bibfield{author}{\bibinfo{person}{Geethu Miriam~Jacob} {and} \bibinfo{person}{Bjorn Stenger}.} \bibinfo{year}{2021}\natexlab{}.
\newblock \showarticletitle{Facial {Action} {Unit} {Detection} {With} {Transformers}}. In \bibinfo{booktitle}{\emph{2021 {IEEE}/{CVF} {Conference} on {Computer} {Vision} and {Pattern} {Recognition} ({CVPR})}}. \bibinfo{publisher}{IEEE}, \bibinfo{address}{Nashville, TN, USA}, \bibinfo{pages}{7676--7685}.
\newblock
\showISBNx{978-1-66544-509-2}
\urldef\tempurl%
\url{https://doi.org/10.1109/CVPR46437.2021.00759}
\showDOI{\tempurl}


\bibitem[Nie et~al\mbox{.}(2020)]%
        {nie_spiders_2020}
\bibfield{author}{\bibinfo{person}{Jingping Nie}, \bibinfo{person}{Yigong Hu}, \bibinfo{person}{Yuanyuting Wang}, \bibinfo{person}{Stephen Xia}, {and} \bibinfo{person}{Xiaofan Jiang}.} \bibinfo{year}{2020}\natexlab{}.
\newblock \showarticletitle{{SPIDERS}: {Low}-{Cost} {Wireless} {Glasses} for {Continuous} {In}-{Situ} {Bio}-{Signal} {Acquisition} and {Emotion} {Recognition}}. In \bibinfo{booktitle}{\emph{2020 {IEEE}/{ACM} {Fifth} {International} {Conference} on {Internet}-of-{Things} {Design} and {Implementation} ({IoTDI})}}. \bibinfo{pages}{27--39}.
\newblock
\urldef\tempurl%
\url{https://doi.org/10.1109/IoTDI49375.2020.00011}
\showDOI{\tempurl}


\bibitem[Nie et~al\mbox{.}(2023)]%
        {nie_time_2023}
\bibfield{author}{\bibinfo{person}{Yuqi Nie}, \bibinfo{person}{Nam~H. Nguyen}, \bibinfo{person}{Phanwadee Sinthong}, {and} \bibinfo{person}{Jayant Kalagnanam}.} \bibinfo{year}{2023}\natexlab{}.
\newblock \bibinfo{title}{A {Time} {Series} is {Worth} 64 {Words}: {Long}-term {Forecasting} with {Transformers}}.
\newblock
\newblock
\urldef\tempurl%
\url{http://arxiv.org/abs/2211.14730}
\showURL{%
\tempurl}
\newblock
\shownote{arXiv:2211.14730 [cs]}.


\bibitem[Park et~al\mbox{.}(2012)]%
        {park_online_2012}
\bibfield{author}{\bibinfo{person}{Jun-geun Park}, \bibinfo{person}{Ami Patel}, \bibinfo{person}{Dorothy Curtis}, \bibinfo{person}{Seth Teller}, {and} \bibinfo{person}{Jonathan Ledlie}.} \bibinfo{year}{2012}\natexlab{}.
\newblock \showarticletitle{Online pose classification and walking speed estimation using handheld devices}. In \bibinfo{booktitle}{\emph{Proceedings of the 2012 {ACM} {Conference} on {Ubiquitous} {Computing}}} \emph{(\bibinfo{series}{{UbiComp} '12})}. \bibinfo{publisher}{Association for Computing Machinery}, \bibinfo{address}{New York, NY, USA}, \bibinfo{pages}{113--122}.
\newblock
\showISBNx{978-1-4503-1224-0}
\urldef\tempurl%
\url{https://doi.org/10.1145/2370216.2370235}
\showDOI{\tempurl}


\bibitem[Qian et~al\mbox{.}(2019)]%
        {qian_novel_2019}
\bibfield{author}{\bibinfo{person}{Hangwei Qian}, \bibinfo{person}{Sinno~Jialin Pan}, \bibinfo{person}{Bingshui Da}, {and} \bibinfo{person}{Chunyan Miao}.} \bibinfo{year}{2019}\natexlab{}.
\newblock \showarticletitle{A {Novel} {Distribution}-{Embedded} {Neural} {Network} for {Sensor}-{Based} {Activity} {Recognition}}.
\newblock  (\bibinfo{year}{2019}), \bibinfo{pages}{5614--5620}.
\newblock
\urldef\tempurl%
\url{https://www.ijcai.org/proceedings/2019/779}
\showURL{%
\tempurl}


\bibitem[Qian et~al\mbox{.}(2022)]%
        {qian_what_2022}
\bibfield{author}{\bibinfo{person}{Hangwei Qian}, \bibinfo{person}{Tian Tian}, {and} \bibinfo{person}{Chunyan Miao}.} \bibinfo{year}{2022}\natexlab{}.
\newblock \showarticletitle{What {Makes} {Good} {Contrastive} {Learning} on {Small}-{Scale} {Wearable}-based {Tasks}?}. In \bibinfo{booktitle}{\emph{Proceedings of the 28th {ACM} {SIGKDD} {Conference} on {Knowledge} {Discovery} and {Data} {Mining}}} \emph{(\bibinfo{series}{{KDD} '22})}. \bibinfo{publisher}{Association for Computing Machinery}, \bibinfo{address}{New York, NY, USA}, \bibinfo{pages}{3761--3771}.
\newblock
\showISBNx{978-1-4503-9385-0}
\urldef\tempurl%
\url{https://doi.org/10.1145/3534678.3539134}
\showDOI{\tempurl}


\bibitem[Richardson et~al\mbox{.}(2017)]%
        {richardson_learning_2017}
\bibfield{author}{\bibinfo{person}{Elad Richardson}, \bibinfo{person}{Matan Sela}, \bibinfo{person}{Roy Or-El}, {and} \bibinfo{person}{Ron Kimmel}.} \bibinfo{year}{2017}\natexlab{}.
\newblock \showarticletitle{Learning {Detailed} {Face} {Reconstruction} from a {Single} {Image}}. In \bibinfo{booktitle}{\emph{2017 {IEEE} {Conference} on {Computer} {Vision} and {Pattern} {Recognition} ({CVPR})}}. \bibinfo{publisher}{IEEE}, \bibinfo{address}{Honolulu, HI}, \bibinfo{pages}{5553--5562}.
\newblock
\showISBNx{978-1-5386-0457-1}
\urldef\tempurl%
\url{https://doi.org/10.1109/CVPR.2017.589}
\showDOI{\tempurl}


\bibitem[Roter et~al\mbox{.}(2006)]%
        {roter_expression_2006}
\bibfield{author}{\bibinfo{person}{Debra~L. Roter}, \bibinfo{person}{Richard~M. Frankel}, \bibinfo{person}{Judith~A. Hall}, {and} \bibinfo{person}{David Sluyter}.} \bibinfo{year}{2006}\natexlab{}.
\newblock \showarticletitle{The expression of emotion through nonverbal behavior in medical visits. {Mechanisms} and outcomes}.
\newblock \bibinfo{journal}{\emph{Journal of General Internal Medicine}} \bibinfo{volume}{21 Suppl 1}, \bibinfo{number}{Suppl 1} (\bibinfo{date}{Jan.} \bibinfo{year}{2006}), \bibinfo{pages}{S28--34}.
\newblock
\showISSN{1525-1497}
\urldef\tempurl%
\url{https://doi.org/10.1111/j.1525-1497.2006.00306.x}
\showDOI{\tempurl}


\bibitem[Roth et~al\mbox{.}({[n.\,d.]})]%
        {roth_adaptive_nodate}
\bibfield{author}{\bibinfo{person}{Joseph Roth}, \bibinfo{person}{Yiying Tong}, {and} \bibinfo{person}{Xiaoming Liu}.} \bibinfo{year}{[n.\,d.]}\natexlab{}.
\newblock \showarticletitle{Adaptive {3D} {Face} {Reconstruction} {From} {Unconstrained} {Photo} {Collections}}.
\newblock  (\bibinfo{year}{[n.\,d.]}).
\newblock


\bibitem[Roth et~al\mbox{.}(2015)]%
        {roth_unconstrained_2015}
\bibfield{author}{\bibinfo{person}{Joseph Roth}, \bibinfo{person}{Yiying Tong}, {and} \bibinfo{person}{Xiaoming Liu}.} \bibinfo{year}{2015}\natexlab{}.
\newblock \showarticletitle{Unconstrained {3D} face reconstruction}. In \bibinfo{booktitle}{\emph{2015 {IEEE} {Conference} on {Computer} {Vision} and {Pattern} {Recognition} ({CVPR})}}. \bibinfo{pages}{2606--2615}.
\newblock
\urldef\tempurl%
\url{https://doi.org/10.1109/CVPR.2015.7298876}
\showDOI{\tempurl}
\newblock
\shownote{ISSN: 1063-6919}.


\bibitem[Saphala et~al\mbox{.}(2022)]%
        {saphala_proximity-based_2022}
\bibfield{author}{\bibinfo{person}{Addythia Saphala}, \bibinfo{person}{Rui Zhang}, {and} \bibinfo{person}{Oliver Amft}.} \bibinfo{year}{2022}\natexlab{}.
\newblock \showarticletitle{Proximity-based {Eating} {Event} {Detection} in {Smart} {Eyeglasses} with {Expert} and {Data} {Models}}. In \bibinfo{booktitle}{\emph{Proceedings of the 2022 {ACM} {International} {Symposium} on {Wearable} {Computers}}} \emph{(\bibinfo{series}{{ISWC} '22})}. \bibinfo{publisher}{Association for Computing Machinery}, \bibinfo{address}{New York, NY, USA}, \bibinfo{pages}{59--63}.
\newblock
\showISBNx{978-1-4503-9424-6}
\urldef\tempurl%
\url{https://doi.org/10.1145/3544794.3558476}
\showDOI{\tempurl}


\bibitem[Scheirer et~al\mbox{.}(1999)]%
        {scheirer_expression_1999}
\bibfield{author}{\bibinfo{person}{Jocelyn Scheirer}, \bibinfo{person}{Raul Fernandez}, {and} \bibinfo{person}{Rosalind~W. Picard}.} \bibinfo{year}{1999}\natexlab{}.
\newblock \showarticletitle{Expression glasses: a wearable device for facial expression recognition}. In \bibinfo{booktitle}{\emph{{CHI} '99 {Extended} {Abstracts} on {Human} {Factors} in {Computing} {Systems}}} \emph{(\bibinfo{series}{{CHI} {EA} '99})}. \bibinfo{publisher}{Association for Computing Machinery}, \bibinfo{address}{New York, NY, USA}, \bibinfo{pages}{262--263}.
\newblock
\showISBNx{978-1-58113-158-1}
\urldef\tempurl%
\url{https://doi.org/10.1145/632716.632878}
\showDOI{\tempurl}


\bibitem[Song et~al\mbox{.}(2022)]%
        {song_facelistener_2022}
\bibfield{author}{\bibinfo{person}{Xingzhe Song}, \bibinfo{person}{Kai Huang}, {and} \bibinfo{person}{Wei Gao}.} \bibinfo{year}{2022}\natexlab{}.
\newblock \showarticletitle{{FaceListener}: {Recognizing} {Human} {Facial} {Expressions} via {Acoustic} {Sensing} on {Commodity} {Headphones}}. In \bibinfo{booktitle}{\emph{2022 21st {ACM}/{IEEE} {International} {Conference} on {Information} {Processing} in {Sensor} {Networks} ({IPSN})}}. \bibinfo{pages}{145--157}.
\newblock
\urldef\tempurl%
\url{https://doi.org/10.1109/IPSN54338.2022.00019}
\showDOI{\tempurl}


\bibitem[Tong et~al\mbox{.}(2007)]%
        {tong_facial_2007}
\bibfield{author}{\bibinfo{person}{Yan Tong}, \bibinfo{person}{Wenhui Liao}, {and} \bibinfo{person}{Qiang Ji}.} \bibinfo{year}{2007}\natexlab{}.
\newblock \showarticletitle{Facial {Action} {Unit} {Recognition} by {Exploiting} {Their} {Dynamic} and {Semantic} {Relationships}}.
\newblock \bibinfo{journal}{\emph{IEEE Transactions on Pattern Analysis and Machine Intelligence}} \bibinfo{volume}{29}, \bibinfo{number}{10} (\bibinfo{date}{Oct.} \bibinfo{year}{2007}), \bibinfo{pages}{1683--1699}.
\newblock
\showISSN{0162-8828}
\urldef\tempurl%
\url{https://doi.org/10.1109/TPAMI.2007.1094}
\showDOI{\tempurl}


\bibitem[Trivedi et~al\mbox{.}(2021)]%
        {trivedi_wifimod_2021}
\bibfield{author}{\bibinfo{person}{Amee Trivedi}, \bibinfo{person}{Kate Silverstein}, \bibinfo{person}{Emma Strubell}, \bibinfo{person}{Prashant Shenoy}, {and} \bibinfo{person}{Mohit Iyyer}.} \bibinfo{year}{2021}\natexlab{}.
\newblock \showarticletitle{{WiFiMod}: {Transformer}-based {Indoor} {Human} {Mobility} {Modeling} using {Passive} {Sensing}}. In \bibinfo{booktitle}{\emph{Proceedings of the 4th {ACM} {SIGCAS} {Conference} on {Computing} and {Sustainable} {Societies}}} \emph{(\bibinfo{series}{{COMPASS} '21})}. \bibinfo{publisher}{Association for Computing Machinery}, \bibinfo{address}{New York, NY, USA}, \bibinfo{pages}{126--137}.
\newblock
\showISBNx{978-1-4503-8453-7}
\urldef\tempurl%
\url{https://doi.org/10.1145/3460112.3471951}
\showDOI{\tempurl}


\bibitem[Uchida et~al\mbox{.}(2018)]%
        {uchida_identification_2018}
\bibfield{author}{\bibinfo{person}{Marco~C. Uchida}, \bibinfo{person}{Renato Carvalho}, \bibinfo{person}{Vitor~Daniel Tessutti}, \bibinfo{person}{Reury Frank~Pereira Bacurau}, \bibinfo{person}{Hélio~José Coelho-Júnior}, \bibinfo{person}{Luciane~Portas Capelo}, \bibinfo{person}{Heloiza~Prando Ramos}, \bibinfo{person}{Marcia~Calixto Dos~Santos}, \bibinfo{person}{Luís Felipe~Milano Teixeira}, {and} \bibinfo{person}{Paulo~Henrique Marchetti}.} \bibinfo{year}{2018}\natexlab{}.
\newblock \showarticletitle{Identification of muscle fatigue by tracking facial expressions}.
\newblock \bibinfo{journal}{\emph{PloS One}} \bibinfo{volume}{13}, \bibinfo{number}{12} (\bibinfo{year}{2018}), \bibinfo{pages}{e0208834}.
\newblock
\showISSN{1932-6203}
\urldef\tempurl%
\url{https://doi.org/10.1371/journal.pone.0208834}
\showDOI{\tempurl}


\bibitem[Uema and Inoue(2017)]%
        {uema_jins_2017}
\bibfield{author}{\bibinfo{person}{Yuji Uema} {and} \bibinfo{person}{Kazutaka Inoue}.} \bibinfo{year}{2017}\natexlab{}.
\newblock \showarticletitle{{JINS} {MEME} algorithm for estimation and tracking of concentration of users}. In \bibinfo{booktitle}{\emph{Proceedings of the 2017 {ACM} {International} {Joint} {Conference} on {Pervasive} and {Ubiquitous} {Computing} and {Proceedings} of the 2017 {ACM} {International} {Symposium} on {Wearable} {Computers}}}. \bibinfo{publisher}{ACM}, \bibinfo{address}{Maui Hawaii}, \bibinfo{pages}{297--300}.
\newblock
\showISBNx{978-1-4503-5190-4}
\urldef\tempurl%
\url{https://doi.org/10.1145/3123024.3123189}
\showDOI{\tempurl}


\bibitem[Ullah and Kim(2024)]%
        {ullah_optimized_2024}
\bibfield{author}{\bibinfo{person}{Shan Ullah} {and} \bibinfo{person}{Deok-Hwan Kim}.} \bibinfo{year}{2024}\natexlab{}.
\newblock \showarticletitle{An {Optimized} {EMG} {Encoder} to {Minimize} {Soft} {Speech} {Loss} for {Speech} to {EMG} {Conversions}}. \bibinfo{publisher}{IEEE Computer Society}, \bibinfo{pages}{215--218}.
\newblock
\showISBNx{9798350370027}
\urldef\tempurl%
\url{https://doi.org/10.1109/BigComp60711.2024.00041}
\showDOI{\tempurl}


\bibitem[Valle et~al\mbox{.}(2019)]%
        {valle_face_2019}
\bibfield{author}{\bibinfo{person}{Roberto Valle}, \bibinfo{person}{José~M. Buenaposada}, \bibinfo{person}{Antonio Valdés}, {and} \bibinfo{person}{Luis Baumela}.} \bibinfo{year}{2019}\natexlab{}.
\newblock \showarticletitle{Face alignment using a {3D} deeply-initialized ensemble of regression trees}.
\newblock \bibinfo{journal}{\emph{Computer Vision and Image Understanding}}  \bibinfo{volume}{189} (\bibinfo{date}{Dec.} \bibinfo{year}{2019}), \bibinfo{pages}{102846}.
\newblock
\showISSN{1077-3142}
\urldef\tempurl%
\url{https://doi.org/10.1016/j.cviu.2019.102846}
\showDOI{\tempurl}


\bibitem[Vazquez-Rodriguez et~al\mbox{.}(2022)]%
        {vazquez-rodriguez_transformer-based_2022}
\bibfield{author}{\bibinfo{person}{Juan Vazquez-Rodriguez}, \bibinfo{person}{Grégoire Lefebvre}, \bibinfo{person}{Julien Cumin}, {and} \bibinfo{person}{James~L. Crowley}.} \bibinfo{year}{2022}\natexlab{}.
\newblock \showarticletitle{Transformer-{Based} {Self}-{Supervised} {Learning} for {Emotion} {Recognition}}. In \bibinfo{booktitle}{\emph{2022 26th {International} {Conference} on {Pattern} {Recognition} ({ICPR})}}. \bibinfo{pages}{2605--2612}.
\newblock
\urldef\tempurl%
\url{https://doi.org/10.1109/ICPR56361.2022.9956027}
\showDOI{\tempurl}
\newblock
\shownote{ISSN: 2831-7475}.


\bibitem[Verma et~al\mbox{.}(2021)]%
        {verma_expressear_2021}
\bibfield{author}{\bibinfo{person}{Dhruv Verma}, \bibinfo{person}{Sejal Bhalla}, \bibinfo{person}{Dhruv Sahnan}, \bibinfo{person}{Jainendra Shukla}, {and} \bibinfo{person}{Aman Parnami}.} \bibinfo{year}{2021}\natexlab{}.
\newblock \showarticletitle{{ExpressEar}: {Sensing} {Fine}-{Grained} {Facial} {Expressions} with {Earables}}.
\newblock \bibinfo{journal}{\emph{Proceedings of the ACM on Interactive, Mobile, Wearable and Ubiquitous Technologies}} \bibinfo{volume}{5}, \bibinfo{number}{3} (\bibinfo{date}{Sept.} \bibinfo{year}{2021}), \bibinfo{pages}{1--28}.
\newblock
\showISSN{2474-9567}
\urldef\tempurl%
\url{https://doi.org/10.1145/3478085}
\showDOI{\tempurl}


\bibitem[Wan et~al\mbox{.}(2023)]%
        {wan_eegformer_2023}
\bibfield{author}{\bibinfo{person}{Zhijiang Wan}, \bibinfo{person}{Manyu Li}, \bibinfo{person}{Shichang Liu}, \bibinfo{person}{Jiajin Huang}, \bibinfo{person}{Hai Tan}, {and} \bibinfo{person}{Wenfeng Duan}.} \bibinfo{year}{2023}\natexlab{}.
\newblock \showarticletitle{{EEGformer}: {A} transformer–based brain activity classification method using {EEG} signal}.
\newblock \bibinfo{journal}{\emph{Frontiers in Neuroscience}}  \bibinfo{volume}{17} (\bibinfo{date}{March} \bibinfo{year}{2023}).
\newblock
\showISSN{1662-453X}
\urldef\tempurl%
\url{https://doi.org/10.3389/fnins.2023.1148855}
\showDOI{\tempurl}
\newblock
\shownote{Publisher: Frontiers}.


\bibitem[Weise et~al\mbox{.}(2009)]%
        {weise_faceoff_2009}
\bibfield{author}{\bibinfo{person}{Thibaut Weise}, \bibinfo{person}{Hao Li}, \bibinfo{person}{Luc Van~Gool}, {and} \bibinfo{person}{Mark Pauly}.} \bibinfo{year}{2009}\natexlab{}.
\newblock \showarticletitle{Face/{Off}: live facial puppetry}. In \bibinfo{booktitle}{\emph{Proceedings of the 2009 {ACM} {SIGGRAPH}/{Eurographics} {Symposium} on {Computer} {Animation}}} \emph{(\bibinfo{series}{{SCA} '09})}. \bibinfo{publisher}{Association for Computing Machinery}, \bibinfo{address}{New York, NY, USA}, \bibinfo{pages}{7--16}.
\newblock
\showISBNx{978-1-60558-610-6}
\urldef\tempurl%
\url{https://doi.org/10.1145/1599470.1599472}
\showDOI{\tempurl}


\bibitem[Wu et~al\mbox{.}(2013)]%
        {wu_modelling_2013}
\bibfield{author}{\bibinfo{person}{Tim Wu}, \bibinfo{person}{Alice P.~L. Hung}, \bibinfo{person}{Peter Hunter}, {and} \bibinfo{person}{Kumar Mithraratne}.} \bibinfo{year}{2013}\natexlab{}.
\newblock \showarticletitle{Modelling facial expressions: {A} framework for simulating nonlinear soft tissue deformations using embedded {3D} muscles}.
\newblock \bibinfo{journal}{\emph{Finite Elements in Analysis and Design}}  \bibinfo{volume}{76} (\bibinfo{date}{Nov.} \bibinfo{year}{2013}), \bibinfo{pages}{63--70}.
\newblock
\showISSN{0168-874X}
\urldef\tempurl%
\url{https://doi.org/10.1016/j.finel.2013.08.002}
\showDOI{\tempurl}


\bibitem[Wu et~al\mbox{.}(2021)]%
        {wu_bioface-3d_2021}
\bibfield{author}{\bibinfo{person}{Yi Wu}, \bibinfo{person}{Vimal Kakaraparthi}, \bibinfo{person}{Zhuohang Li}, \bibinfo{person}{Tien Pham}, \bibinfo{person}{Jian Liu}, {and} \bibinfo{person}{Phuc Nguyen}.} \bibinfo{year}{2021}\natexlab{}.
\newblock \showarticletitle{{BioFace}-{3D}: continuous 3d facial reconstruction through lightweight single-ear biosensors}. In \bibinfo{booktitle}{\emph{Proceedings of the 27th {Annual} {International} {Conference} on {Mobile} {Computing} and {Networking}}}. \bibinfo{publisher}{ACM}, \bibinfo{address}{New Orleans Louisiana}, \bibinfo{pages}{350--363}.
\newblock
\showISBNx{978-1-4503-8342-4}
\urldef\tempurl%
\url{https://doi.org/10.1145/3447993.3483252}
\showDOI{\tempurl}


\bibitem[Xie et~al\mbox{.}(2023)]%
        {xie_mm3dface_2023}
\bibfield{author}{\bibinfo{person}{Jiahong Xie}, \bibinfo{person}{Hao Kong}, \bibinfo{person}{Jiadi Yu}, \bibinfo{person}{Yingying Chen}, \bibinfo{person}{Linghe Kong}, \bibinfo{person}{Yanmin Zhu}, {and} \bibinfo{person}{Feilong Tang}.} \bibinfo{year}{2023}\natexlab{}.
\newblock \showarticletitle{{mm3DFace}: {Nonintrusive} {3D} {Facial} {Reconstruction} {Leveraging} {mmWave} {Signals}}. In \bibinfo{booktitle}{\emph{Proceedings of the 21st {Annual} {International} {Conference} on {Mobile} {Systems}, {Applications} and {Services}}}. \bibinfo{publisher}{ACM}, \bibinfo{address}{Helsinki Finland}, \bibinfo{pages}{462--474}.
\newblock
\showISBNx{9798400701108}
\urldef\tempurl%
\url{https://doi.org/10.1145/3581791.3596839}
\showDOI{\tempurl}


\bibitem[Xie et~al\mbox{.}(2021)]%
        {xie_acoustic-based_2021}
\bibfield{author}{\bibinfo{person}{Wentao Xie}, \bibinfo{person}{Qian Zhang}, {and} \bibinfo{person}{Jin Zhang}.} \bibinfo{year}{2021}\natexlab{}.
\newblock \showarticletitle{Acoustic-based {Upper} {Facial} {Action} {Recognition} for {Smart} {Eyewear}}.
\newblock \bibinfo{journal}{\emph{Proceedings of the ACM on Interactive, Mobile, Wearable and Ubiquitous Technologies}} \bibinfo{volume}{5}, \bibinfo{number}{2} (\bibinfo{year}{2021}), \bibinfo{pages}{41:1--41:28}.
\newblock
\urldef\tempurl%
\url{https://doi.org/10.1145/3448105}
\showDOI{\tempurl}


\bibitem[Yan et~al\mbox{.}(2020)]%
        {yan_frownonerror_2020}
\bibfield{author}{\bibinfo{person}{Yukang Yan}, \bibinfo{person}{Chun Yu}, \bibinfo{person}{Wengrui Zheng}, \bibinfo{person}{Ruining Tang}, \bibinfo{person}{Xuhai Xu}, {and} \bibinfo{person}{Yuanchun Shi}.} \bibinfo{year}{2020}\natexlab{}.
\newblock \showarticletitle{{FrownOnError}: {Interrupting} {Responses} from {Smart} {Speakers} by {Facial} {Expressions}}. In \bibinfo{booktitle}{\emph{Proceedings of the 2020 {CHI} {Conference} on {Human} {Factors} in {Computing} {Systems}}}. \bibinfo{publisher}{ACM}, \bibinfo{address}{Honolulu HI USA}, \bibinfo{pages}{1--14}.
\newblock
\showISBNx{978-1-4503-6708-0}
\urldef\tempurl%
\url{https://doi.org/10.1145/3313831.3376810}
\showDOI{\tempurl}


\bibitem[Zarins(2018)]%
        {zarins_anatomy_2018}
\bibfield{author}{\bibinfo{person}{Uldis Zarins}.} \bibinfo{year}{2018}\natexlab{}.
\newblock \bibinfo{booktitle}{\emph{Anatomy of {Facial} {Expression}}}.
\newblock \bibinfo{publisher}{Exonicus Incorporated}.
\newblock
\showISBNx{978-0-9903411-1-6}
\newblock
\shownote{Google-Books-ID: CqMyngAACAAJ}.


\bibitem[Zeng et~al\mbox{.}(2018)]%
        {zeng_understanding_2018}
\bibfield{author}{\bibinfo{person}{Ming Zeng}, \bibinfo{person}{Haoxiang Gao}, \bibinfo{person}{Tong Yu}, \bibinfo{person}{Ole~J. Mengshoel}, \bibinfo{person}{Helge Langseth}, \bibinfo{person}{Ian Lane}, {and} \bibinfo{person}{Xiaobing Liu}.} \bibinfo{year}{2018}\natexlab{}.
\newblock \bibinfo{title}{Understanding and {Improving} {Recurrent} {Networks} for {Human} {Activity} {Recognition} by {Continuous} {Attention}}.
\newblock
\newblock
\urldef\tempurl%
\url{https://doi.org/10.48550/arXiv.1810.04038}
\showDOI{\tempurl}
\newblock
\shownote{arXiv:1810.04038 [cs, stat]}.


\bibitem[Zhang et~al\mbox{.}(2023b)]%
        {zhang_i_2023}
\bibfield{author}{\bibinfo{person}{Shijia Zhang}, \bibinfo{person}{Taiting Lu}, \bibinfo{person}{Hao Zhou}, \bibinfo{person}{Yilin Liu}, \bibinfo{person}{Runze Liu}, {and} \bibinfo{person}{Mahanth Gowda}.} \bibinfo{year}{2023}\natexlab{b}.
\newblock \showarticletitle{I {Am} an {Earphone} and {I} {Can} {Hear} {My} {User}’s {Face}: {Facial} {Landmark} {Tracking} {Using} {Smart} {Earphones}}.
\newblock \bibinfo{journal}{\emph{ACM Transactions on Internet of Things}} \bibinfo{volume}{5}, \bibinfo{number}{1} (\bibinfo{year}{2023}), \bibinfo{pages}{1:1--1:29}.
\newblock
\urldef\tempurl%
\url{https://doi.org/10.1145/3614438}
\showDOI{\tempurl}


\bibitem[Zhang et~al\mbox{.}(2018)]%
        {zhang_dense_2018}
\bibfield{author}{\bibinfo{person}{Shu Zhang}, \bibinfo{person}{Hui Yu}, \bibinfo{person}{Ting Wang}, \bibinfo{person}{Lin Qi}, \bibinfo{person}{Junyu Dong}, {and} \bibinfo{person}{Honghai Liu}.} \bibinfo{year}{2018}\natexlab{}.
\newblock \showarticletitle{Dense {3D} facial reconstruction from a single depth image in unconstrained environment}.
\newblock \bibinfo{journal}{\emph{Virtual Reality}} \bibinfo{volume}{22}, \bibinfo{number}{1} (\bibinfo{date}{March} \bibinfo{year}{2018}), \bibinfo{pages}{37--46}.
\newblock
\showISSN{1434-9957}
\urldef\tempurl%
\url{https://doi.org/10.1007/s10055-017-0311-6}
\showDOI{\tempurl}


\bibitem[Zhang et~al\mbox{.}(2023a)]%
        {zhang_bleselect_2023}
\bibfield{author}{\bibinfo{person}{Tengxiang Zhang}, \bibinfo{person}{Zitong Lan}, \bibinfo{person}{Chenren Xu}, \bibinfo{person}{Yanrong Li}, {and} \bibinfo{person}{Yiqiang Chen}.} \bibinfo{year}{2023}\natexlab{a}.
\newblock \showarticletitle{{BLEselect}: {Gestural} {IoT} {Device} {Selection} via {Bluetooth} {Angle} of {Arrival} {Estimation} from {Smart} {Glasses}}.
\newblock \bibinfo{journal}{\emph{Proceedings of the ACM on Interactive, Mobile, Wearable and Ubiquitous Technologies}} \bibinfo{volume}{6}, \bibinfo{number}{4} (\bibinfo{date}{Jan.} \bibinfo{year}{2023}), \bibinfo{pages}{198:1--198:28}.
\newblock
\urldef\tempurl%
\url{https://doi.org/10.1145/3569482}
\showDOI{\tempurl}


\bibitem[Zhang et~al\mbox{.}(2019)]%
        {zhang_facilitating_2019}
\bibfield{author}{\bibinfo{person}{Tengxiang Zhang}, \bibinfo{person}{Xin Yi}, \bibinfo{person}{Ruolin Wang}, \bibinfo{person}{Jiayuan Gao}, \bibinfo{person}{Yuntao Wang}, \bibinfo{person}{Chun Yu}, \bibinfo{person}{Simin Li}, {and} \bibinfo{person}{Yuanchun Shi}.} \bibinfo{year}{2019}\natexlab{}.
\newblock \showarticletitle{Facilitating {Temporal} {Synchronous} {Target} {Selection} through {User} {Behavior} {Modeling}}.
\newblock \bibinfo{journal}{\emph{Proceedings of the ACM on Interactive, Mobile, Wearable and Ubiquitous Technologies}} \bibinfo{volume}{3}, \bibinfo{number}{4} (\bibinfo{date}{Dec.} \bibinfo{year}{2019}), \bibinfo{pages}{1--24}.
\newblock
\showISSN{24749567}
\urldef\tempurl%
\url{https://doi.org/10.1145/3369839}
\showDOI{\tempurl}


\bibitem[Zhang et~al\mbox{.}(2023c)]%
        {zhang_mmfer_2023}
\bibfield{author}{\bibinfo{person}{Xi Zhang}, \bibinfo{person}{Yu Zhang}, \bibinfo{person}{Zhenguo Shi}, {and} \bibinfo{person}{Tao Gu}.} \bibinfo{year}{2023}\natexlab{c}.
\newblock \showarticletitle{{mmFER}: {Millimetre}-wave {Radar} based {Facial} {Expression} {Recognition} for {Multimedia} {IoT} {Applications}}.
\newblock In \bibinfo{booktitle}{\emph{Proceedings of the 29th {Annual} {International} {Conference} on {Mobile} {Computing} and {Networking}}}. Number~23. \bibinfo{publisher}{Association for Computing Machinery}, \bibinfo{address}{New York, NY, USA}, \bibinfo{pages}{1--15}.
\newblock
\showISBNx{978-1-4503-9990-6}
\urldef\tempurl%
\url{https://doi.org/10.1145/3570361.3592515}
\showURL{%
\tempurl}


\bibitem[Zhang et~al\mbox{.}(2022)]%
        {zhang_if-convtransformer_2022}
\bibfield{author}{\bibinfo{person}{Ye Zhang}, \bibinfo{person}{Longguang Wang}, \bibinfo{person}{Huiling Chen}, \bibinfo{person}{Aosheng Tian}, \bibinfo{person}{Shilin Zhou}, {and} \bibinfo{person}{Yulan Guo}.} \bibinfo{year}{2022}\natexlab{}.
\newblock \showarticletitle{{IF}-{ConvTransformer}: {A} {Framework} for {Human} {Activity} {Recognition} {Using} {IMU} {Fusion} and {ConvTransformer}}.
\newblock \bibinfo{journal}{\emph{Proceedings of the ACM on Interactive, Mobile, Wearable and Ubiquitous Technologies}} \bibinfo{volume}{6}, \bibinfo{number}{2} (\bibinfo{date}{July} \bibinfo{year}{2022}), \bibinfo{pages}{1--26}.
\newblock
\showISSN{2474-9567}
\urldef\tempurl%
\url{https://doi.org/10.1145/3534584}
\showDOI{\tempurl}


\end{thebibliography}

\appendix

\end{document}